\documentclass[structabstract]{aa}  
\usepackage{graphicx}
\usepackage{epstopdf}
\usepackage{color}
\usepackage{longtable}

\usepackage{txfonts}
\usepackage{natbib}
\def\gr{$\gamma$-ray}

\begin{document}
   \title{High Galactic latitude {\it Fermi} sources of $\gamma$-rays with energies above 100~GeV}

    \author{A. Neronov
          \inst{1},
          D.Semikoz\inst{2,3}
          \and
          Ie.Vovk
          \inst{1}
          }

   \institute{ISDC Data Centre for Astrophysics, Ch. d'Ecogia 16, 1290, Versoix, Switzerland \\
              \email{Andrii.Neronov@unige.ch}
         \and
             APC, 10 rue Alice Domon et Leonie Duquet, F-75205 Paris Cedex 13, France \and
Institute for Nuclear Research RAS, 60th October Anniversary prosp. 7a, Moscow, 117312, Russia
\\
             \email{Dmitri Semikoz <dmitri.semikoz@apc.univ-paris7.fr>}
             }

%   \date{Received ; accepted }

% \abstract{}{}{}{}{} 
% 5 {} token are mandatory
 
  \abstract
   {}
   {We present a catalog of sources of very high energy ($E>100$~GeV) \gr s detected by the Large Area Telescope (LAT) onboard the \textit{ Fermi} Gamma-Ray Space Telescope at Galactic latitudes $|b|\ge 10^\circ$.}
   {We cross-correlate the directions of individual photons with energies above 100~GeV detected by \textit{ Fermi} /LAT  with the first year \textit{ Fermi} catalog of sources detected at lower energies. We find a significant correlation between the arrival directions of the highest energy photons and positions of \textit{ Fermi} sources, with the possibility of chance coincidences at the level of  $10^{-45}$. We present a list of \textit{ Fermi} sources contributing to the correlation signal.  A similar analysis is performed to cross-correlate the catalog of BL Lac objects with the highest energy photons detected by {\it Fermi}. }
   {We produce a catalog of high Galactic latitude \textit{ Fermi} sources visible at 100-300 ~GeV energies. The catalog is divided into two parts. The first part contains a list of 50 higher significance sources among which there can be three possible false detections.  The second part of the catalog contains a list of 25 lower significance sources, among which six are possibly false detections. 
   Finally we identify eight additional sources from the cross-correlation analysis with the BL Lac catalog. The reported sources of $E>100$~GeV \gr s span a broad range of redshifts, up to $z\sim 1$. Most of the sources are BL Lac type objects. Only 17 out of 83 objects in our list were previously reported as VHE \gr\ sources.}
   {}


   \maketitle

%%%%%%%%%%%%%%%%%%%%%%%%%%%%%%%%%%%%%
\section{Introduction}
%%%%%%%%%%%%%%%%%%%%%%%%%%%%%%%%%%%%%

Ground-based Cherenkov \gr\ telescopes HESS, MAGIC, and VERITAS have discovered a population of sources of very high energy (VHE) ($E\ge 100$~GeV) \gr s. Except for the sources discovered in the Galactic plane survey by HESS \citep{HESS_survey_science,HESS_survey}, most of the sources were discovered in dedicated pointed observations.  Surveys of large regions of the VHE \gr\ sky with the existing  Cherenkov telescopes are difficult because of the too-narrow size of the field of view. The wide field-of-view ground-based \gr\ detectors MILAGRO \citep{milagro} and Tibet \citep{tibet} arrays have produced a systematic survey of the VHE \gr\ sky. The energy threshold of the air shower arrays such as MILAGRO and TIBET is  rather high (in the multi-TeV band) so that only sources with spectra extending well above 1 TeV could be detected.

All-sky monitoring of \gr\ sources at the energies $E\sim 100$~GeV became possible with the start of operation of the Large Area Telescope (LAT) onboard the {\it Fermi} Gamma-Ray Space Telescope.  Compared to the ground-based \gr\ telescopes, the LAT has a much smaller effective area ($\sim 1$~m$^2$, compared to $\sim 10^5$~m$^2$ for the ground-based \gr\ telescopes). At the same time, at the energies above 100~GeV the {\it Fermi} signal is almost background-free (in contrast to the ground based telescopes in which the signal has to be identified on top of the strong background created by cosmic rays and optical/UV night sky background).  \citet{paper1} searched for  the point sources  of $E\ge 100$~GeV \gr s  using   {\it Fermi} data
and found eight significant excesses with at least three photons within a $0.1^\circ$ circle corresponding to the 68\% containment radius of the point spread function (PSF) of the LAT. Seven excesses were 
associated with known VHE \gr\ sources, while the remaining one was identified with the head-tail radio galaxy IC 310. Detection of IC 310 in the VHE band was later confirmed by the MAGIC telescope \citep{IC310_MAGIC}. Owing to the moderate collection area of {\it Fermi}, only the brightest VHE \gr\ sources were detected individually in the {\it Fermi} VHE \gr\ sky survey. All other known VHE \gr\ sources at Galactic latitudes $|b|\ge 10^\circ$ gave $\le 2$ photons within the LAT PSF  circle and could not be found  from the analysis of the data above $100$~GeV alone. 

A complementary method for identifying the sources of $E\ge 100$~GeV photons detected by {\it Fermi} is to use  prior knowledge of source positions on the sky and to verify which of the already known sources could have produced the highest energy  \gr s detected by the LAT telescope. In other words, sources of $E\ge 100$~GeV \gr s could be identified also via cross-correlation of arrival directions of the $E\ge 100$~GeV \gr s with the source positions on the sky. This approach to the identification of the sources was previously applied to the analysis of EGRET data above $10$~GeV by \citet{dingus01} and \citet{10GeV_EGRET}. 

We perform the cross-correlation analysis of the arrival directions of \gr s with energies above 100~GeV detected by {\it Fermi} at Galactic latitudes $|b|\ge 10^\circ$ with  the first year  {\it Fermi}  source catalog \citep{fermi_catalog}. Our analysis results in a catalog of 50 high Galactic latitude sources that correlate with the arrival directions of $100~\mbox{GeV} \le E \le 300$~GeV \gr s within a 68\% containment circle of LAT PSF. Among these 50 sources, seven sources that correlate with $\ge 3$ VHE \gr s (considered previously by \citet{paper1}) are detected above 100 GeV with significance $\ge 8\sigma$. Six sources that correlate with two photons are detected with a significance of around $4 - 5\sigma$.  Each individual source that correlates with just one VHE \gr\ has a significance of around $3\sigma$ in this energy band, the overall significance of the detection of the entire source set being very high. The chance coincidence probability of $E>100$~GeV $\gamma$-ray arrival directions correlating with the source positions is $< 10^{-40}$ (excluding the seven high significance sources from \citet{paper1}), which corresponds to the significance of detection of the entire set of $50-7=43$ sources  $>13 \sigma$. The sources correlating with only one $E>100$~GeV photon should be considered as "VHE source candidates". A simple analysis indicates that most of the sources contributing to the correlation signal are real VHE \gr\ sources, only three of them being expected to be false detections. Taking into account that most of the sources in the list are BL Lac type objects, we extend our cross-correlation analysis to the catalog of BL Lacs \citep{veron13} and find eight more sources that correlate with the arrival directions of $100~\mbox{GeV} \le E \le 300$~GeV \gr s and are not listed in the first year {\it Fermi} catalog. For completeness, we list sources from the first year {\it Fermi} catalog for which the $E\ge 100$~GeV \gr s are found within a circle of the radius at which the correlation signal is strongest. Its radius is somewhat larger than the 68\% containment circle of LAT PSF. There are 25 such sources, six of them being expected to be false detections because of the chance coincidence of the arrival direction of the VHE \gr\ with the source position. For energies $E> 300$ GeV, we found correlations with four known sources, one possible new source and one false detection. 

The plan of the paper is as follows. In Section \ref{sec:data} we discuss data selection and data analysis methods. In Section \ref{sec:correlation}, we present the results of the correlation analysis of the arrival directions of $E\ge 100$~GeV \gr s with the sources of the first year {\it Fermi} catalog. In Section \ref{sec:catalog}, we provide the list of sources contributing to the correlation signal. In Section \ref{sec:BL} we perform the correlation analysis with the BL Lac catalog of \citet{veron13} and give the list of additional BL Lacs correlating with the highest energy {\it Fermi} photons, but not present in the first year {\it Fermi} catalog. In Section \ref{sec:300GeV} we discuss correlations of $E\ge 300$~GeV photons with all possible sources discussed above. In Section \ref{sec:individual} we comment upon individual sources. Finally, in Section \ref{sec:discussion} we discuss the results.

%%%%%%%%%%%%%%%%%%%%%%%%%%%%%%%%%%%%%
\section{Data selection and data analysis}
\label{sec:data}
%%%%%%%%%%%%%%%%%%%%%%%%%%%%%%%%%%%%%

For our analysis, we used the LAT data collected in the period between August 4, 2008 and June 25, 2010. The data were filtered using the  {\it gtselect} tool provided by Fermi Science Tools\footnote{http://fermi.gsfc.nasa.gov/ssc/data/analysis/}, so that only \gr\ events ({\tt evcls=3}) with energies above 100~GeV were retained in the analysis. 
The resulting list of photons has 6376 events. 
A fraction of the $E\ge 100$~GeV photons comes from directions close to the Galactic plane, which is a source of significant diffuse \gr\ emission even at photon energies higher than 100~GeV. Presence of strong diffuse emission complicates the analysis of the point source contribution. Taking this into account, we consider only photons coming from Galactic latitudes $|b|\ge 10^\circ$. The cut on the Galactic latitude leaves 4086 photons for the analysis.

For energies $E<300$ GeV, photons are thought to be clearly distinguishable from cosmic rays in {\it Fermi}, while 
at $E>300$  GeV ({\tt evcls=3})  the sample contains an unknown number of misidentified cosmic rays ~\citep{Fermi_300}. We therefore divide data into two energy bins divided at $E=300$ GeV.
 There are 3186 photons with energies $100~\mbox{GeV} \le E \le 300$ GeV, for which we perform our main analysis in the following sections, and there are in addition 900 photons with energies $E > 300$ GeV,
 which we discuss separately in  Section~\ref{sec:300GeV}\footnote{For the method discussed below, it does not matter whether the background consists of photons or misidentified cosmic rays. We are grateful to I.Tkachev and P.Tinyakov, who pointed out that we can use photons with $E>300$ GeV in our analysis.}.

The set of photons considered in the analysis includes the list of photons studied by \citet{paper1}.  In our analysis, we exclude the high-confidence sources of VHE \gr s identified by  \citet{paper1}. We also exclude 75 \gr\ photons associated with these eight sources from the {\it Fermi} $E\ge 100$~GeV photon list, i.e. found   within 0.2 degrees of those sources. Otherwise, these sources would dominate the correlation signal.  Nevertheless, we include  the high-confidence sources in the final source list given in the Table \ref{table1}, to obtain a complete catalog of sources of VHE \gr s found by {\it Fermi}. The final photon list used in the analysis includes 3111 photons  with $|b|\ge 10^\circ$.

Photons detected by {\it Fermi}/LAT are divided into two types -- \textit{front-} and \textit{back-}converting. Photons that pair-convert in the top 12 layers of the tracker are classified as front-converting, and are otherwise back-converting \citep{LAT_high_latit_survey}. At the same energy, front-converting photons have a somewhat sharper PSF\footnote{http://www-glast.slac.stanford.edu/software/\\ IS/glast\_lat\_performance.htm}. We take this into account by considering the sets of 1179 front- and 1934 back- converted photons separately in our correlation analysis. 

If the sky region around the source is not too crowded, the point spread functions of individual {\it Fermi} sources do not overlap at the energies above $\sim 1$~ GeV. At high Galactic latitudes, the strength and complexity of the diffuse Galactic \gr\ background at energies above $1$~GeV is is far lower than the low-latitude / low-energy background. We calculate the spectra of the sources using two methods, one the unbinned likelihood analysis performed using the {\it gtlike} tool in narrow energy bins (see {\tt http://fermi.gsfc.nasa.gov/ssc/data/analysis/ scitools/likelihood\_tutorial.html}) and the other using the aperture photometry method for  the same energy bins. In each case, we verify that the two analysis methods provide consistent results. The aperture photometry method is most useful at the highest energies, at which the photon statistics is low. In this method, we first calculate for each source and each energy  the number of source counts within a circle of the radius equal to the 95\% containment circle of the LAT PSF and calculate the LAT exposure within this circle using the {\it gtexposure} tool. The errors in the measurements are calculated based on a Poisson distribution, which provides the correct description of the data at low photon statistics.  To estimate the diffuse sky background at the position of the source, we calculate the number of background counts in a ring with the outer radius $3^\circ$ and the inner radius equal to the 95\% containment radius of the LAT PSF, centered on the source.  We verified that the results obtained using such an intuitively simple spectral extraction procedure are consistent with those obtained via spectral extraction using likelihood analysis in narrow energy bins and/or likelihood analysis in the wide energy range using a specific broad-band spectral model.

%%%%%%%%%%%%%%%%%%%%%%%%%%%%%%%%%%%%%
\section{Correlation between arrival directions of $E\ge 100$~GeV photons and sources from the first year {\it Fermi} catalog}
\label{sec:correlation}
%%%%%%%%%%%%%%%%%%%%%%%%%%%%%%%%%%%%%

%%%%%%%%%%%%%%%%%%%%%%%%%%%%%%%%%%%%%

\begin{figure}[htbp]
\begin{center}
\includegraphics[width=0.7\linewidth,angle=-90]{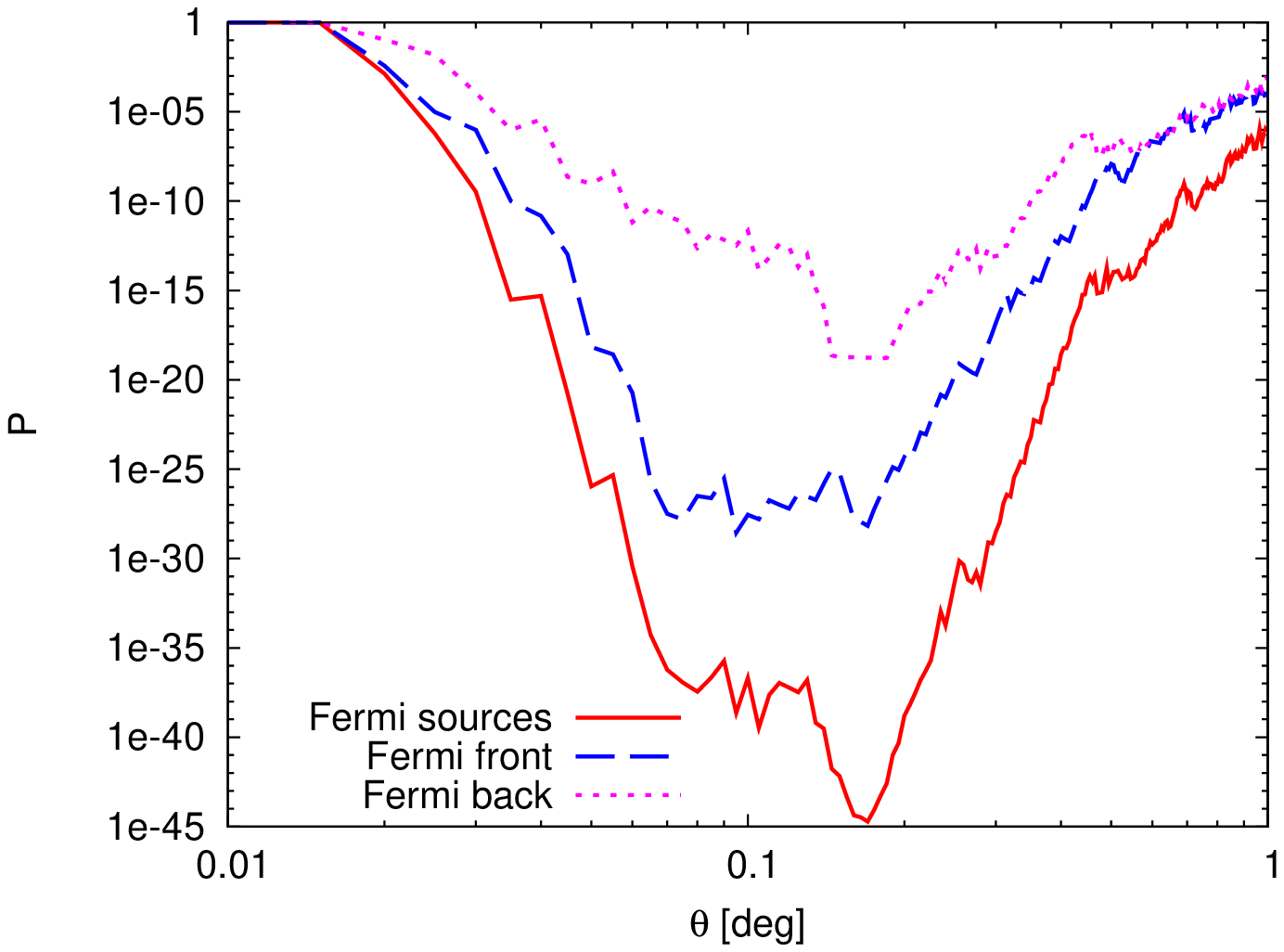}
\end{center}
\caption{Dependence on the chance probability of correlation between arrival directions of $E\ge 100$~GeV photons and positions of {\it Fermi} sources as a function of the radius of the search circle. Solid curve corresponds to all photons. Dashed and dotted curves correspond to, respectively, front- and back-converted photons.}
\label{fig:probability}
\end{figure}
%%%%%%%%%%%%%%%%%%%%%%%%%%%%%%%%%%%%%

To find sources of $E\ge 100$~GeV \gr s that cannot individually be recognized in the {\it Fermi} data in this energy band,
we apply a method similar to the one discussed by \citet{dingus01} and \citet{10GeV_EGRET}. Using this method, sources producing just a few photons could be identified.  A fraction of identified sources might be false detections because of a chance coincidence of the source position with the arrival direction of a background photon. The fraction of positive-to-false detections could be readily estimated. Real sources should largely outnumber the false source detections. This makes the list of sources contributing to the correlation a valuable "input catalog" for the observations with ground-based \gr\ telescopes that  are more sensitive than {\it Fermi} in the VHE \gr\ band, but, in contrast to {\it Fermi}, do not have all-sky survey capabilities.

As a first choice, we took the first year {\it Fermi}  catalog as an input catalog for the correlation analysis \citep{fermi_catalog}. The catalog contains 1451 sources, 1043 of which are objects with $|b|\ge$10 degrees. Removing  seven confirmed VHE \gr\ sources from the list of \citet{paper1}  (IC 310 is not in the {\it Fermi} catalog), we obtain an input catalog of $N_{\rm source}=1036$ sources.
%%%%%%%%%%%%%%%%%%%%%%%%%%%%%%%%%%%%%

\begin{figure}[htbp]
\begin{center}
\includegraphics[width=0.7\linewidth,angle=-90]{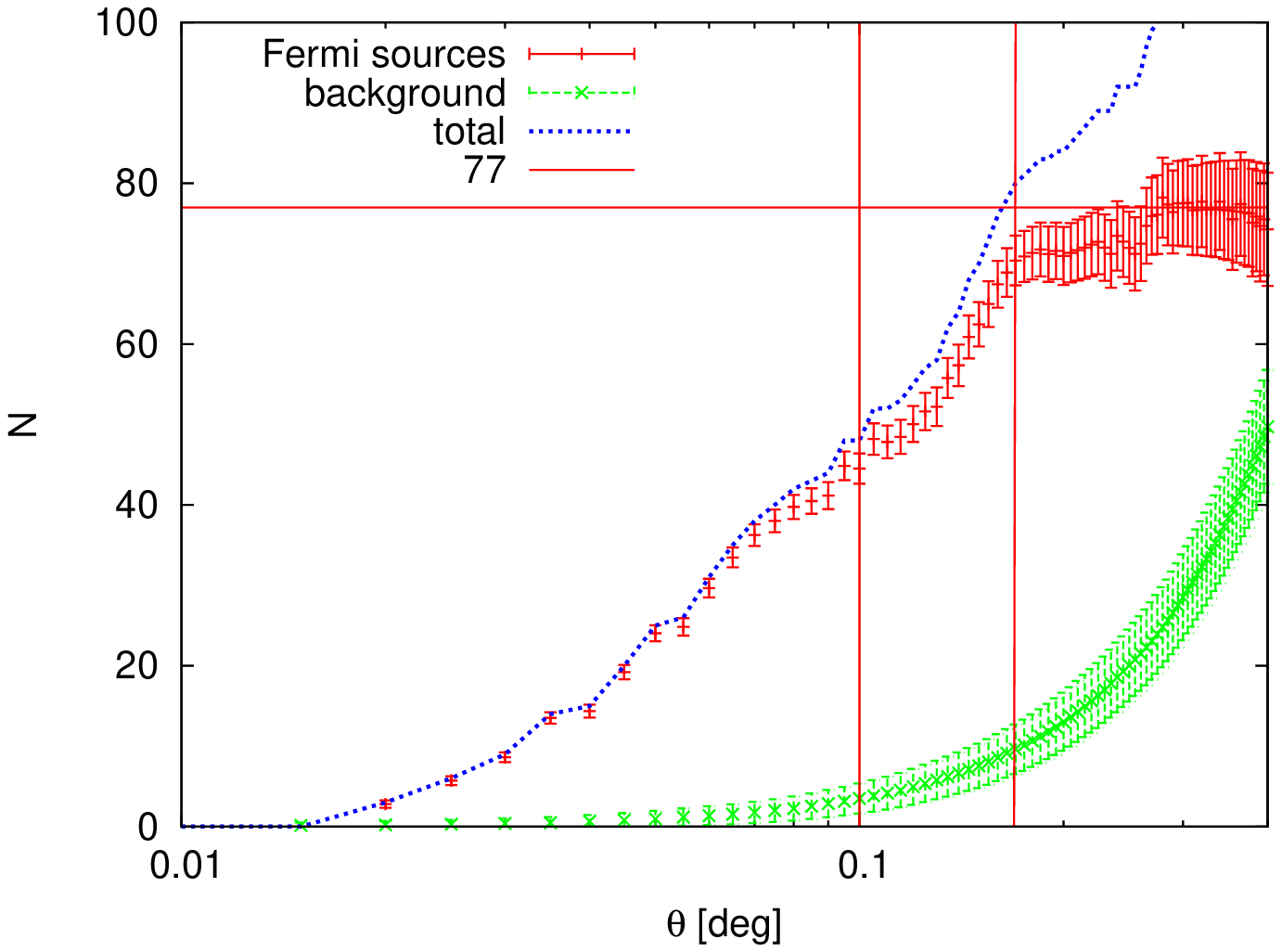}
\end{center}
\caption{Number of events that come from  {\it Fermi} sources as a function of the radius of the search circle (red data points). Error bars indicate fluctuations in the background when determining the signal from the sources.
Average number of background photons is shown with green line. Vertical line at 0.1 degree is the radius of  the 68 \% containment circle of {\it Fermi} PSF. The vertical line
at 0.17 degree shows the  position of the dip in the probability shown in 
Fig.~\ref{fig:probability}.  The horizontal line indicates the  asymptotic number of photons from all the sources.}
\label{fig:npart}
\end{figure}
%%%%%%%%%%%%%%%%%%%%%%%%%%%%%%%%%%%%%

The probability that a \gr\  photon originates within a given distance $\theta$ from one source in the catalog is estimated as the ratio of the area of the circle of radius $\theta$ to the part of the sky at $|b|\ge 10^\circ$
\begin{equation}
p_1=\frac{\pi\theta^2}{4\pi(1-\cos(80^\circ))}\simeq 9.2 \cdot 10^{-7} \left[\frac{\theta}{0.1^\circ}\right]^2.
\end{equation}
The total  probability of a photon originating within an  angle $\theta$ of one of the catalog  sources is then 
\begin{equation}
p=N_{\rm source} \cdot p_1 = 9.5 \cdot 10^{-4} \left[\frac{N_{\rm source}}{1036}\right]\left[\frac{\theta}{0.1^\circ}\right]^2.
\end{equation}  
The probability that $K$ or more photons from $N_{\gamma} =  3111 $ \gr s  come within the angle $\theta$ form any of sources in the catalog by chance is given by the binomial probability
\begin{equation}
P(\theta ) = \sum_{k=K}^{N_{tot}} p^k (1-p)^{N_{\gamma}-k}\frac{N_{\gamma}!}{(N_{\gamma}-k)! k!}.
\label{binom}
\end{equation}
In Fig.~\ref{fig:probability}, we plot this probability as a function of angle $\theta$. The probability that \gr s with energies higher than $100$~GeV come from directions close to the positions of {\it Fermi} sources by chance is $P< 10^{-45}$, i.e. the chance coincidence hypothesis is firmly ruled out.  

The function $P(\theta)$ has two minima at around 0.1 degrees and 0.17 degrees. This is explained by how precision with which the arrival direction is determined for the front and back  converted \gr s differs. To verify this,  we divide the whole photon list into two parts corresponding to the front and back converted \gr s and plot the contributions of those photons separately. From Fig. \ref{fig:probability}, one can see that the minimum at 0.1 degree is due to front photons, while the minimum at 0.17 degrees is due to both front and back photons.

In Fig. \ref{fig:npart},  we plot  the number of source photons with arrival directions within the angle $\theta$ from the catalog of Fermi sources and compare it to the expected number of background events within the same distance from the sources. To estimate the number of background events, we used two alternative methods. First, we estimated the expected background near each source counting all photons within 10 degrees from it
and estimating the fraction of photons that originate within the angle  $\theta$. This method can be used only for $\theta\lesssim 0.5^\circ$. For $\theta\gtrsim 0.5^\circ$, circles around the {\it Fermi} sources cover a significant fraction of the sky and overlap.   In the second method, we estimated the background by generating 
3111 photons in the sky outside the Galactic plane $|b|>10^\circ$. We followed the {\it Fermi} exposure which had been estimated using the {\it gtexposure} tool by  keeping Galactic b-values of 
events and generating randomly l-values. This takes into account the remaining contribution of the Galaxy for $|b|>10^\circ$. 
After that, we average over many  Monte Carlo simulations. Both methods gave similar results for small angles.

Vertical lines in Fig. \ref{fig:npart} show the angular positions of the two minima of probability seen in Fig.~\ref{fig:probability}.  The minima correspond to the places
where the steep rise in the number of signal events changes to a plateau. The asymptote of the distribution of signal counts at large $\theta$ gives an estimate of the total number of \gr s contributing to the correlation.  
One can see that only $\sim 77$ out of  3111 events come from  known {\it Fermi} sources. 48 for these 77 events are displaced by less than $0.1^\circ$ from the catalog source position. In this circle, one expects only  $N_{B,\ 0.1}\simeq 3$ background events. No more than seven background photons are expected at the 95\% confidence level and no more than nine background photons are expected at the 99.5\% confidence level.
The number of events contributing to the correlation within  $\theta=0.17^\circ$ circle is $N_{S,\ 0.17}\simeq 71$, while the expected number of background events is  $N_{B,\ 0.17}\simeq 7.7$ (13 at the 95\% confidence level, 15 at the 99\% confidence level).

%%%%%%%%%%%%%%%%%%%%%%%%%%%%%%%%%%%%%
\section{Catalog of extragalactic VHE \gr\ sources} 
\label{sec:catalog}
%%%%%%%%%%%%%%%%%%%%%%%%%%%%%%%%%%%%%

The list of sources contributing to the correlation signal within $\theta=0.1^\circ$ is given in Table \ref{table1}. There are $N_{\rm Source,\ 0.1}=50-7=43$ sources, if the sources from the list of \citet{paper1} are excluded.  For $N_\gamma=3111$, only $p\cdot N_\gamma\simeq 3$ photons  on average can be at the distance $\theta\le 0.1^\circ$ from a catalog source by chance. This means that  3 out of 43 new VHE \gr\ sources listed in Table \ref{table1} are expected to be false detections (and no more than 7[9] sources are false detections at the 95\% [99.5\%] confidence level).

%%%%%%%%%%%%%%%%%%%%%%%%%%%%%%%%%%%%%
\begin{table*}
\begin{tabular}{|l|l|l|l|l|l|l|l|l|l|l|l|l|l|}
\hline
&1FGL            &  RA      &  Dec     &  $N_{68}$   &      $P_{68}$       & $N_{68}$     & $N_{min}$  &         $P$         & TS  & Type$^1$ & Name                           & $z$    \\
&                &          &          & $_{30-100}$ &     $_{30-100}$     & $_{100-300}$ & [$N_{95}$] &     $_{100-300}$    &     &          &                                &        \\
\hline 
1  &J0033.5-1921 &  8.3913  & -19.3650 &      6      & $2.1\cdot 10^{-14}$ &     1b       &      1     & $2.3\cdot 10^{-5}$  &  21 &   BL V   & RBS 76                         & 0.61   \\
2  &J0054.9-2455 &  13.7337 & -24.9290 &      0      & --                  &     1f       &            & $3.2\cdot 10^{-3}$  &  18 &          &                                &        \\   
3  &J0110.0-4023 &  17.5250 & -40.3887 &      1      & $1.3\cdot 10^{-2}$  &     1f       &            & $2.4\cdot 10^{-3}$  &  12 &          &                                &        \\
4  &J0209.3-5229 &  32.3392 & -52.4907 &      2      & $1.1\cdot 10^{-4}$  &     1f       &            & $3.5\cdot 10^{-3}$  &  12 &   BL Vf  & RBS 285                        &        \\
5  &J0213.2+2244 &  33.3225 &  22.7489 &      2      & $1.6\cdot  10^{-4}$ &     1b       &            & $2.6\cdot 10^{-3}$  &  13 &   QSO    & 1RXS                           &        \\
   &             &          &          &             &                     &              &            &                     &     &          & J021252.2+22                   &        \\
6  &J0222.6+4302 &  35.6681 &  43.0385 &     16      & $4.6\cdot 10^{-40}$ &     3f1b     &      1     & $1.4\cdot 10^{-13}$ &  75 &   BL V   & {\bf \underline{3C 66A}}       & 0.444  \\
7  &J0237.5-3603 &  39.3788 & -36.0530 &      3      & $2.3\cdot 10^{-7}$  &     1b       &            & $2.4\cdot 10^{-3}$  &  15 &   BL V   & RBS 334                        &        \\
8  &J0303.5-2406 &  45.8890 & -24.1124 &      9      & $2.0\cdot 10^{-23}$ &     2f       &     [1]    & $1.8\cdot 10^{-6}$  &  39 &   BL Vf  & \underline{PKS 0301-243}       & 0.26   \\
9  &J0315.9-2609 &  48.9933 & -26.1581 &      1      & $1.4\cdot 10^{-2}$  &     2b       &            & $3.5\cdot 10^{-6}$  &  28 &   BL V   & \underline{RX J0316.2-2607}    & 0.443  \\
10 &J0316.3-6438 &  49.0996 & -64.6411 &      0      &  --                 &     1b       &            & $1.5\cdot 10^{-3}$  &  13 &          &                                &        \\
11 &J0322.1+2336 &  50.5396 &  23.6108 &      2      & $2.0\cdot 10^{-4}$  &     1f       &            & $4  \cdot 10^{-3}$  &  16 &   BL Vf  & RGB J0321+236                  &        \\
12 &J0325.9-1649 &  51.4780 & -16.8174 &      0      & --                  &     1b       &            & $2.5\cdot 10^{-3}$  &  12 &   BL V   & RBS 421                        & 0.29   \\
13 &J0338.8+1313 &  54.7212 &  13.2314 &      0      &  --                 &     1b       &            & $3.1\cdot 10^{-3}$  &  13 &          &                                &        \\
14 &J0416.8+0107 &  64.2069 &   1.1244 &      0      & --                  &     1f       &            & $2.5\cdot 10^{-3}$  &  14 &   BL Vf  & {\bf 2E 0414+0057}             &        \\
15 &J0428.6-3756 &  67.1567 & -37.9412 &      9      & $3.4\cdot 10^{-22}$ &     1f       &            & $2.9\cdot 10^{-3}$  &  18 &   BL     & PKS 0426-380                   & 1.11   \\
16 &J0505.9+6121 &  76.4853 &  61.3527 &      1      & $1.9\cdot 10^{-2}$  &     1f       &     [2]    & $1.6\cdot 10^{-6}$  &  21 &          &                                &        \\
17 &J0507.9+6738 &  76.9931 &  67.6367 &     12      & $2.7\cdot 10^{-29}$ &     2f2b     &      3     & $3.8\cdot 10^{-19}$ &  96 &   BL V   & {\bf \underline{1ES 0502+675}} & 0.341  \\
18 &J0543.8-5531 &  85.9512 & -55.5301 &      1      & $2.0\cdot 10^{-2}$  &     1f1b     &            & $5.3\cdot 10^{-6}$  &  34 &   BL V   & \underline{RBS 679}            &        \\
19 &J0650.7+2503 & 102.6817 &  25.0622 &      1      & $1.0\cdot 10^{-2}$  &     1f       &            & $2.2\cdot 10^{-3}$  &  15 &   BL V   & 1ES 0647+250                   & 0.203  \\
20 &J0710.6+5911 & 107.6595 &  59.1853 &      3      & $2.3\cdot 10^{-6}$  &     1f       &            & $2.0\cdot 10^{-3}$  &  17 &   BL     & {\bf RX J0710.4+5908}          & 0.125  \\
21 &J0721.9+7120 & 110.4794 &  71.3448 &      9      & $6.3\cdot 10^{-21}$ &     1f       &            & $3.6\cdot 10^{-3}$  &  13 &   BL Vf  & {\bf S5 0716+71}               & 0.3    \\
22 &J0745.2+7438 & 116.3215 &  74.6415 &      0      & --                  &     1b       &            & $3.9\cdot 10^{-3}$  &  10 &   BL V   & 1ES 0737+746                   & 0.315  \\
23 &J0809.4+3455 & 122.3539 &  34.9331 &      0      & --                  &     1f       &            & $2.5\cdot 10^{-3}$  &  14 &   BL Vf  & B2 0806+35                    & 0.0825 \\
24 &J0809.5+5219 & 122.3856 &  52.3178 &      5      & $3.1\cdot 10^{-11}$ &     1f1b     &            & $5.9\cdot 10^{-6}$  &  25 &   BL Vf  & {\bf \underline{1ES 0806+524}}             & 0.138  \\
25 &J0816.4-1311 & 124.1121 & -13.1933 &      5      & $5.5\cdot 10^{-13}$ &     1f       &            & $1.8\cdot 10^{-3}$  &  13 &   BL Vf  & PMN J0816-1311                 &        \\
26 &J0905.5+1356 & 136.3916 &  13.9483 &      1      & $1.5\cdot 10^{-2}$  &     1b       &            & $4.0\cdot 10^{-3}$  &  14 &   QSO?   & SDSS J090513.28                & 1.12   \\
   &             &          &          &             &                     &              &            &                     &     &          & +140240.3                      &        \\
27 &J0915.7+2931 & 138.9412 &  29.5319 &      3      & $8.6\cdot 10^{-7}$  &     1f       &            & $2.7\cdot 10^{-3}$  &  13 &   BL Vf  & B2 0912+29                     &        \\
28 &J0953.0-0838 & 148.2673 &  -8.6461 &      2      & $9.5\cdot 10^{-5}$  &     1f       &            & $2.6\cdot 10^{-3}$  &  15 &   BL Vf  & PMN J0953-840                  &        \\
29 &J0957.7+5523 & 149.4315 &  55.3922 &      6      & $2.1\cdot 10^{-13}$ &     1b       &            & $3.6\cdot 10^{-3}$  &  12 &   QSO    & 4C +55.17                      & 0.8955 \\
30 &J1015.1+4927 & 153.7885 &  49.4526 &      5      & $2.9\cdot 10^{-11}$ &     1b       &     2      & $2.6\cdot 10^{-7}$  &  35 &   BL V   & {\bf \underline{1ES 1011+496}} & 0.212  \\
31 &J1104.4+3812 & 166.1247 &  38.2108 &     57      & 0                   &     14f8b    &     3[1]   & $0$                 & 494 &   BL V   & {\bf \underline{Mkn 421}}      & 0.03   \\
32 &J1117.1+2013 & 169.2993 &  20.2222 &      3      & $5.5\cdot 10^{-7}$  &     1f       &            & $2.9\cdot 10^{-3}$  &  12 &   BL Vf  & RBS 958                        & 0.1    \\
33 &J1133.1+0033 & 173.2792 &  0.56306 &      0      & --                  &     1f       &            & $2.9\cdot 10^{-3}$  &  11 &   BL V   & PKS B1130+008                  & 1.223  \\
34 &J1217.7+3007 & 184.4463 &  30.1206 &      2      & $2.0\cdot 10^{-4}$  &     1b       &            & $3.0\cdot 10^{-3}$  &   8 &   BL V   & B2 1215+30                     & 0.13   \\
35 &J1224.7+2121 & 186.1987 &  21.3633 &      6      & $7.3\cdot 10^{-14}$ &     1f1b     &            & $1.7\cdot 10^{-6}$  &  31 &   FSRQ   & \underline{4C +21.35}          & 0.432  \\
36 &J1309.5+4304 & 197.3933 &  43.0669 &      3      & $8.8\cdot 10^{-7}$  &     1f       &            & $2.9\cdot 10^{-3}$  &  13 &   BL Vf  & B3 1307+433                    & 0.69   \\
37 &J1337.7-1255 & 204.4336 & -12.9289 &      0      & --                  &     1f       &            & $1.5\cdot 10^{-3}$  &  15 &   FSRQ   & PKS 1335-127                   & 0.539  \\
38 &J1437.0+5640 & 219.2617 &  56.6720 &      4      & $5.7\cdot 10^{-9}$  &     1f       &            & $3.4\cdot 10^{-3}$  &  17 &   BL     & RBS 1409                       & 0.15   \\
39 &J1555.7+1111 & 238.9383 &  11.1921 &     25      & 0                   &     3f2b     &     1[1]   & $3  \cdot 10^{-16}$ & 124 &   BL V   & {\bf \underline{PG 1553+113}}  &        \\
40 &J1653.9+3945 & 253.4897 &  39.7527 &     17      & $1.1\cdot 10^{-44}$ &     9f1b     &     1      & $1.5\cdot 10^{-32}$ & 214 &   BL V   & {\bf \underline{Mkn 501}}      & 0.0336 \\
41 &J1722.5+1012 & 260.6494 &  10.2144 &      0      & --                  &     1f       &            & $2.3\cdot 10^{-3}$  &  17 &   FSRQ   & TXS 1720+102                   &        \\
42 &J1744.2+1934 & 266.0710 &  19.5706 &      1      & $2.1\cdot 10^{-2}$  &     1b       &            & $2.5\cdot 10^{-3}$  &  13 &   BL V   & 1ES 1741+196                   & 0.083  \\
43 &J1824.6+1013 & 276.1739 &  10.2180 &      0      &  --                 &     1f       &            & $1.9\cdot 10^{-3}$  &  10 &          &                                &        \\
44 &J2000.0+6508 & 300.0215 &  65.1334 &      7      & $2.1\cdot 10^{-15}$ &     1f       &     [1]    & $1.9\cdot 10^{-4}$  &  20 &   BL Vf  & {\bf 1ES 1959+650}             & 0.048  \\
45 &J2000.9-1749 & 300.2467 & -17.8286 &      1      & $2.2\cdot 10^{-2}$  &     1b       &            & $4.0\cdot 10^{-3}$  &  11 &   FSRQ   & PKS 1958-179                   & 0.65   \\
46 &J2004.8+7004 & 301.2074 &  70.0710 &      0      &  --                 &     1b       &            & $4.3\cdot 10^{-3}$  &  13 &          &                                &        \\
47 &J2009.5-4849 & 302.3761 & -48.8295 &      7      & $1.3\cdot 10^{-15}$ &     4f1b     &     [1]    & $7.5\cdot 10^{-15}$ &  82 &   BL V   & {\bf \underline{PKS 2005-489}} & 0.071  \\
48 &J2158.8-3013 & 329.7117 & -30.2176 &     23      & 0                   &     5f3b     &     1[1]   & $2.2\cdot 10^{-24}$ & 164 &   BL V   & {\bf \underline{PKS 2155-304}} & 0.117  \\
49 &J2314.1+1444 & 348.5487 &  14.7463 &      1      & $1.8\cdot 10^{-2}$  &     1f       &            & $3.1\cdot 10^{-3}$  &  13 &   AGN    & SDSS J231405.46                & 1.319  \\
   &             &          &          &             &                     &              &            &                     &     &          & +143904.6                      &        \\
50 &J2322.6+3435 & 350.6540 &  34.5839 &      1      & $2.0\cdot 10^{-2}$  &     1f       &            & $3.4\cdot 10^{-3}$  &  14 &   BL     & TXS 2320+343                   & 0.098  \\
\hline
\end{tabular}

\caption{List of {\it Fermi} sources for which a $100~\mbox{GeV} \le E  \le 300$~GeV photon is found within the 68\% containment circle of radius $0.1^\circ$.} \tablefoot{ first column: {\it Fermi} source ID; 2-nd column: Right Accention; 3-rd column: Declination; 4-th column: number of photons in the 68\% containment circle in 30-100~GeV band; 5-th column: chance coincidence probability to find given number of photons in the 68\% containment circle in 30-100~GeV band; 6-th column: number of photons in the 68\% containment circle in 100-300~GeV band; "f" marks front photons, "b" marks back photons; 7-th column: number of photons in the  $\theta\le 0.17^\circ$ circle around the source or in the 95\% containment circle of the radius $0.3^\circ$ (square brackets); 8-th column: chance probability to find given number of background photons within $0.1^\circ$ and $0.17^\circ\ [0.3^\circ]$ circles around the source position; 9-th column: TS value in 100-300 GeV energy bin, found with our analysis; 10-th column: source type; 11-th column: alternative name of the source; 12-th column: source redshift. Bold marks the sources which have been previously detected in the VHE band by ground-based \gr\ telescopes. Underlined are sources detected with Fermi in 100-300 GeV band with $TS>25$. The updated version of this table is available online at \underline{http://www.isdc.unige.ch/vhe/.}
 \newline  $^1$ BL -- BL Lac type object; QSO -- quasi-stellar object, AGN -- active galactic nucleus, FSRQ -- flat spectrum radio quasar; FR I -- Fanaroff-Riley type I radio galaxy; V -- also found from the correlation analysis with the BL Lac catalog; Vf -- front photon contributes to the correlation signal with the BL Lac catalog.}
\label{table1}
\end{table*}
%%%%%%%%%%%%%%%%%%%%%%%%%%%%%%%%%%%%%

\begin{figure}
\includegraphics[width=\linewidth]{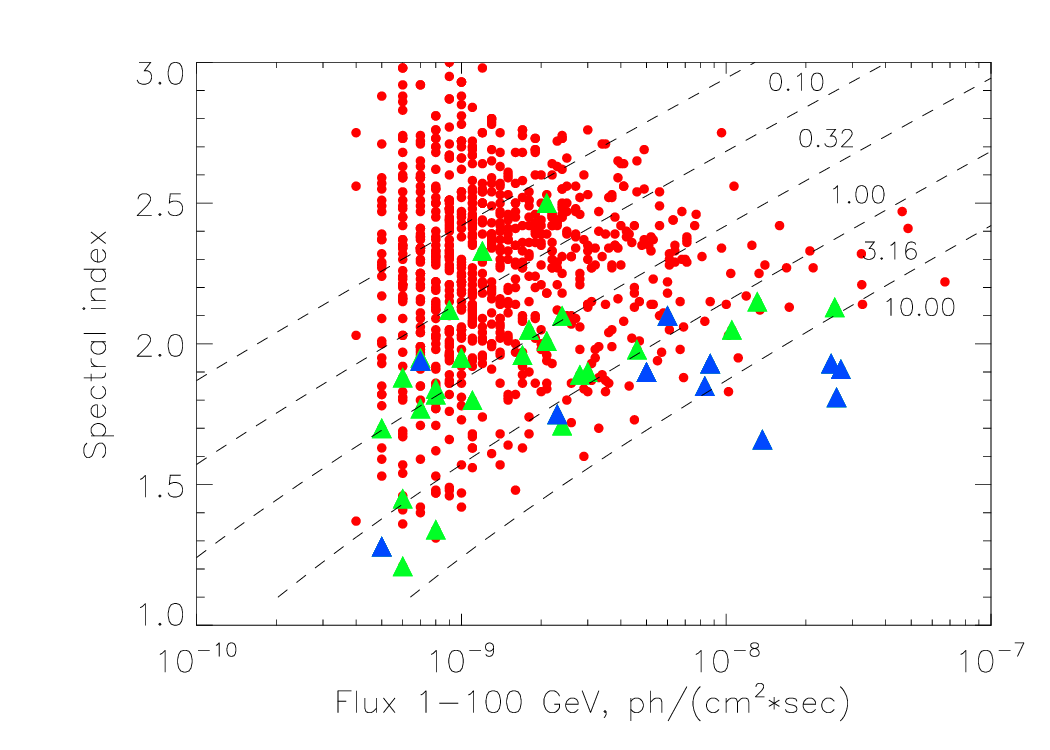}
\caption{Comparison of the distribution of fluxes and photon indices of {\it Fermi} sources of $E\ge 100$~GeV \gr s (triangles) with those of all sources from the {\it Fermi} AGN catalog. Blue triangles mark known TeV \gr\ sources.}
\label{fig:distribution}
\end{figure}

It is not possible to know in advance which of the 43 sources from Table \ref{table1} are false detections. In principle, the ten sources that have more than one photon within the 95\% containment circle of the LAT PSF are not expected to be false detections, because the probability that two background photons come close to one and the same source is much smaller than the same probability for just one background photon.  Additional information that might help us to single out the false detections could be obtained by comparing of the estimate of the source fluxes expected from the extrapolation of the measured spectral characteristics of the sources in the 1-100~GeV energy band with the simple estimates of the flux that produces one photon at $E\ge 100$~GeV within 1.8~year exposure with {\it Fermi}. In principle, a false detection might be spotted if the source contributing to the correlation has low flux and/or a soft spectrum. This is illustrated in Fig. \ref{fig:distribution} where the distribution of  fluxes and photon indices of the AGN from the {\it Fermi} AGN catalog \citep{Fermi_AGN} is compared to the distribution of fluxes and photon indices of the sources from Table \ref{table1}. Dashed lines show the combinations of the spectral parameters that are expected to give a fixed number of photons in the 100-300 GeV band, assuming no high-energy cut-off in the spectrum. 

%%%%%%%%%%%%%%%%%%%%%%%%%%%%%%%%%%%%%
%%%%%%%%%%%%%%%%%%%%%%%%%%%%%%%%%%%%%

\begin{figure*}
\includegraphics[width=\linewidth]{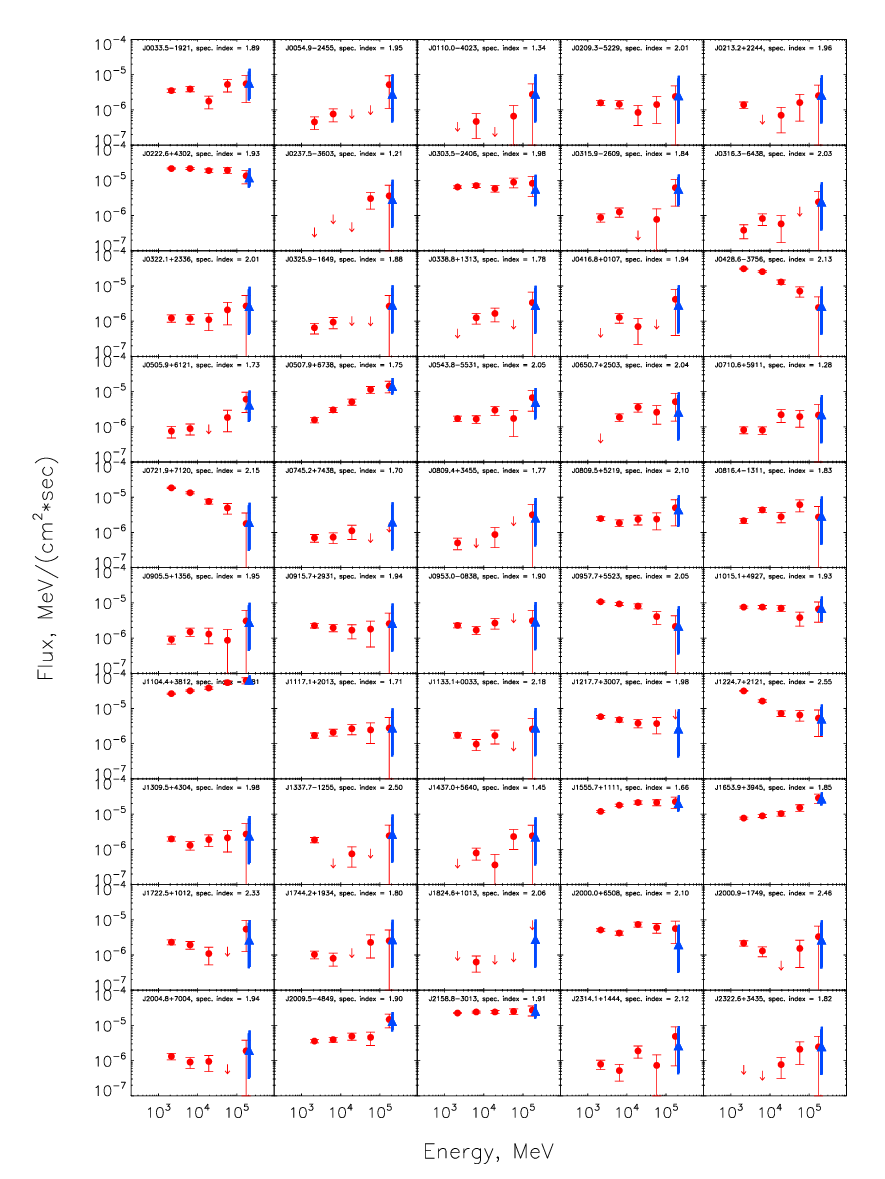}
\caption{ Comparison of the spectra of the sources from Table \ref{table1} in the 1-300~GeV band, obtained with a standard Fermi analysis procedure (red circles represent the flux measurements, red arrows - upper limits), and the estimated source fluxes in the 100-300~GeV band, obtained using aperture photometry technique (blue triangles). Aperture photometry data points in the 100-300 GeV band are artificially shifted to slightly higher energy to ensure that they are visible, and displayed with the measurements from the likelihood analysis in the same energy band.}
\label{fig:all_spectra}
\end{figure*}

Fig. \ref{fig:all_spectra} shows the measurements of the source fluxes in 0.3-1~GeV, 1-3~GeV, 3-10~GeV, and 10-100~GeV for the entire 1.8 yr Fermi/LAT exposure in addition to  the estimates of the source fluxes in the 100-300~GeV band based on the number of $E\ge 100$~GeV detected photons from each source. 

Although the significance of the detection of individual sources of 1(2) photons in the 100-300~GeV band in Table \ref{table1} is around $3\sigma (4\sigma)$, some of the sources are detected with a significance higher than $5\sigma$ in the adjacent 30-100~GeV energy band. The 5-th and 6-th columns of Table \ref{table1} show the numbers of photons and the chance coincidence probabilities derived by comparing the expected number of background photons within $\theta\le 0.12^\circ$ circles (corresponding to the 68\% containment circle at 30~GeV) around the sources with the number of source photons in the 30-100~GeV band. For the sources which are detected with more than $5\sigma$ significance in the 30-100~GeV band we show more detailed spectra above 1~GeV, calculated using all publicly available data of {\it Fermi} collected from August 2008 to June 2010 (rather than the 11 month data used for the analysis of the first year {\it Fermi} catalog).
One can see that for most of the sources the estimates of the source flux in the 100-300~GeV energy bin agree well with the extrapolation of the power-law spectrum from lower energies. 

 %%%%%%%%%%%%%%%%%%%%%%%%%%%%%%%%%%%%%
%%%%%%%%%%%%%%%%%%%%%%%%%%%%%%%%%%%%%

The maximum of the correlation signal (minimum of the chance probability) shown in Fig. \ref{fig:npart} is achieved at the angle $\theta=0.17^\circ$, corresponding to the maximum correlation signal in the back-converted events. The improvement in the correlation achieved with the increase of $\theta$ from $0.1^\circ$ to $0.17^\circ$ means that there are more real events within the rings $0.1\le \theta\le 0.17^\circ$ around their sources, than there are background events in these rings. 

Table \ref{table2} lists the sources contributing to the correlation signal within $\theta\le 0.17^\circ$ but not within $\theta\le 0.1^\circ$. The number of background events in the ring $0.1^\circ\le \theta\le 0.17^\circ$ is estimated to be 6. This implies that 6  out of the 25 additional sources listed in Table \ref{table2} might be false detections. 

%%%%%%%%%%%%%%%%%%%%%%%%%%%%%%%%%%%%%
\begin{table*}
\begin{tabular}{|l|l|ll|c|c|c|c|l|l|l|l|}
\hline
&1FGL &              RA &      Dec	&    $N_{68}$&$P_{68}$     & $N_{min}$      & $P_{min}$     &Type             & Name2  &$z$\\
&&            &    &    $~_{30-100}$        			  &   $~_{30-100}$    & $~_{100-300}$&   $~_{100-300}$           &             &  &\\
\hline 
51 & J0043.6+3424 &  10.9240 &  34.4060 & 1  & $2\cdot 10^{-2}$         &  1f  &$7.7\cdot 10^{-3}$&  & & \\
52 & J0203.5+7234 &  30.8940 &  72.5747 & 0  & --                                 &  1f  &$7.2\cdot 10^{-3}$ & BL  & CGRaBS J0203+7232 & \\
53 & J0440.6+2748 &  70.1738 &  27.8151 &  0 & --                                 &  1f  & $5.2\cdot 10^{-3}$& & B2 0437+27B & \\
54 & J0442.7-0019 &  70.6896 & -0.3192 &  0 & --                                   &  1f  & $7.5\cdot 10^{-3}$& & PKS 0440-00 & \\
55 & J0449.5-4350& 72.3783  &-43.8383 &  12 &  $6.5\cdot 10^{-30}$   &  1f  & $7.8\cdot 10^{-3}$& & PKS 0447-439 & \\
56 & J0536.2-3348 &  84.0605 & -33.8027 & 2  &  $ 1.6 \cdot 10^{-4}$   &  1f  & $8.1\cdot 10^{-3}$& BL  Vf& FRBA J0536-3343 & \\
57 & J0803.1-0339 & 120.7800 &  -3.6544 &  1 & $1.1\cdot 10^{-2}$      &  1f  &$3.1\cdot 10^{-3}$& & & \\
58 & J0856.6+2103 & 134.1578 &  21.0619 & 1 &  $ 1.4 \cdot 10^{-2}$    &  1b & $7.8\cdot 10^{-3}$& FSRQ & OJ 290 &  2.106\\
59 & J0909.2+2310 & 137.3005 &  23.1761 & 1  & $1.7\cdot 10^{-2}$      &  1b &$8.3\cdot 10^{-3}$& BL  V& RX J0909.0+2311    & \\
60 & J1101.3+1009 &  165.3275&  10.1569 & 1  & $1.6\cdot 10^{-2}$       &  1f  &$7.7\cdot 10^{-3}$&   & & \\
61 & J1112.8+3444 & 168.2035 &  34.7363 &  0 & --                                 &  1b & $1.2\cdot 10^{-2}$& FSRQ & CRATES J1112+3446 & 1.9556\\
62 & J1125.5-3559 & 171.3949 & -35.9978 & 0 & --                                  &  2b & $2.5\cdot 10^{-5}$& AGN & CRATES J1125-3557 & \\
63 & J1136.9+2551 & 174.2362 &  25.8602 &  2 & $1.5 \cdot 10^{-4}$      &  1f  & $1.1\cdot 10^{-2}$& BL  V& BZB J1136+2550 & 0.2\\
64 &J1154.0-0008 &  178.5117 & -0.1469  & 0  & --                                  &  1b  &$7.8\cdot 10^{-3}$&   & & \\
65 & J1253.0+5301 & 193.2646 &  53.0245 &  1 &  $1.9 \cdot 10^{-2}$    &  1b  &$8.6\cdot 10^{-3}$& BL  V& CRATES J1253+5301 & \\
66 & J1328.2-4729 & 202.0501 & -47.4991 & 3  & $1.8\cdot 10^{-6}$      &  1f  &$5.6\cdot 10^{-3}$& & & \\
67 & J1345.4+4453 & 206.3740 &  44.8965 & 0  & --                                 &  1b  &$7.7\cdot 10^{-3}$ & FSRQ & B3 1343+451 & \\
68 & J1426.9+2347 & 216.7495 &  23.7987 & 16  &  $7.5\cdot 10^{-43}$ &  2b  & $2.6\cdot 10^{-5}$& BL  V& {\bf PKS 1424+240} &0.16\\
69 & J1428.7+4239 & 217.1790 &  42.6568 &  6 & $1.3\cdot 10^{-14}$    &  1b  &$7.8\cdot 10^{-3}$ & BL  V& {\bf 1ES 1426+428} & 0.129\\
70 & J1553.9+4952 & 238.4886 &  49.8724 & 2  & $1.3\cdot 10^{-4}$      &  1b  &$9.8\cdot 10^{-3}$ && & \\
71 & J1923.5-2104 & 290.8849 & -21.0771 &  0 & --                                &  1f  &$9.4\cdot 10^{-3}$& FSRQ & OV -235 & 0.874\\
72 & J2014.5-0047 &303.6400  &  -0.7858 & 1  &  $2 \cdot 10^{-2}$        &  1b  &$7.9\cdot 10^{-3}$&   & & \\
73 & J2250.1+3825 & 342.5275 &  38.4328 & 3  &  $1.6\cdot 10^{-6}$     &  1b & $7.3\cdot 10^{-3}$& BL V&  3 2247+381 & 0.1187\\
74 &J2323.5+4211 &  350.8912 &  42.1947   &  5 & $2.1 \cdot 10^{-11} $                            &  1b   & $7.\cdot 10^{-3}$& BL  V&  1ES 2321+419 & \\
75 & J2325.8-4043 & 351.4613 & -40.7184 & 1  & $1.4\cdot 10^{-2}$      & 1b  & $8.\cdot 10^{-3}$& BL  V& RXS J23247-4040 & \\
\hline
\end{tabular}

\caption{List of {\it Fermi} sources for which a $100~\mbox{GeV} < E  < 300$~GeV  photon is found within the radius $\theta=0.17^\circ$, but not within $\theta=0.1^\circ$. Notations are the same as in Table 1.}
\label{table2} 
\end{table*}
%%%%%%%%%%%%%%%%%%%%%%%%%%%%%%%%%%%%%

We note that only 16 out of 75  sources listed in Tables \ref{table1} and \ref{table2} are already known sources of VHE \gr s detected by the ground-based \gr\ telescopes. The majority of the sources from Tables \ref{table1} and \ref{table2} are new real sources in the VHE band. Taking into account  that only 9 out of the 75 sources listed in Tables \ref{table1} and \ref{table2} are possible false detections, the {\it Fermi} all sky survey at the energy $E_\gamma\ge 100$~GeV reveals 75-9-16=50 new extragalactic VHE \gr\ sources. This doubles the number of already known extragalactic VHE \gr\ sources \footnote{See http://tevcat.uchicago.edu/}.

Another important point is that the $E\ge 100$~GeV \gr\ sources listed in Tables \ref{table1} and \ref{table2} are distributed over a broad range of redshifts. The "record" redshift of the VHE \gr\ source  so far was $z\simeq 0.5$ for the possible detection of 3C 279 during a short several hour flare by MAGIC \citep{3c279_magic}. In our Table \ref{table1}, several sources have redshifts higher than 0.5.

%%%%%%%%%%%%%%%%%%%%%%%%%%%%%%%%%%%%%
\section{Correlation between arrival directions of $100~\mbox{GeV} \le E  \le 300$~GeV photons and sources from the 13th Veron catalog} 
\label{sec:BL}
%%%%%%%%%%%%%%%%%%%%%%%%%%%%%%%%%%%%%

%%%%%%%%%%%%%%%%%%%%%%%%%%%%%%%%%%%%%

\begin{figure}[htbp]
\begin{center}
\includegraphics[width=0.7\linewidth,angle=-90]{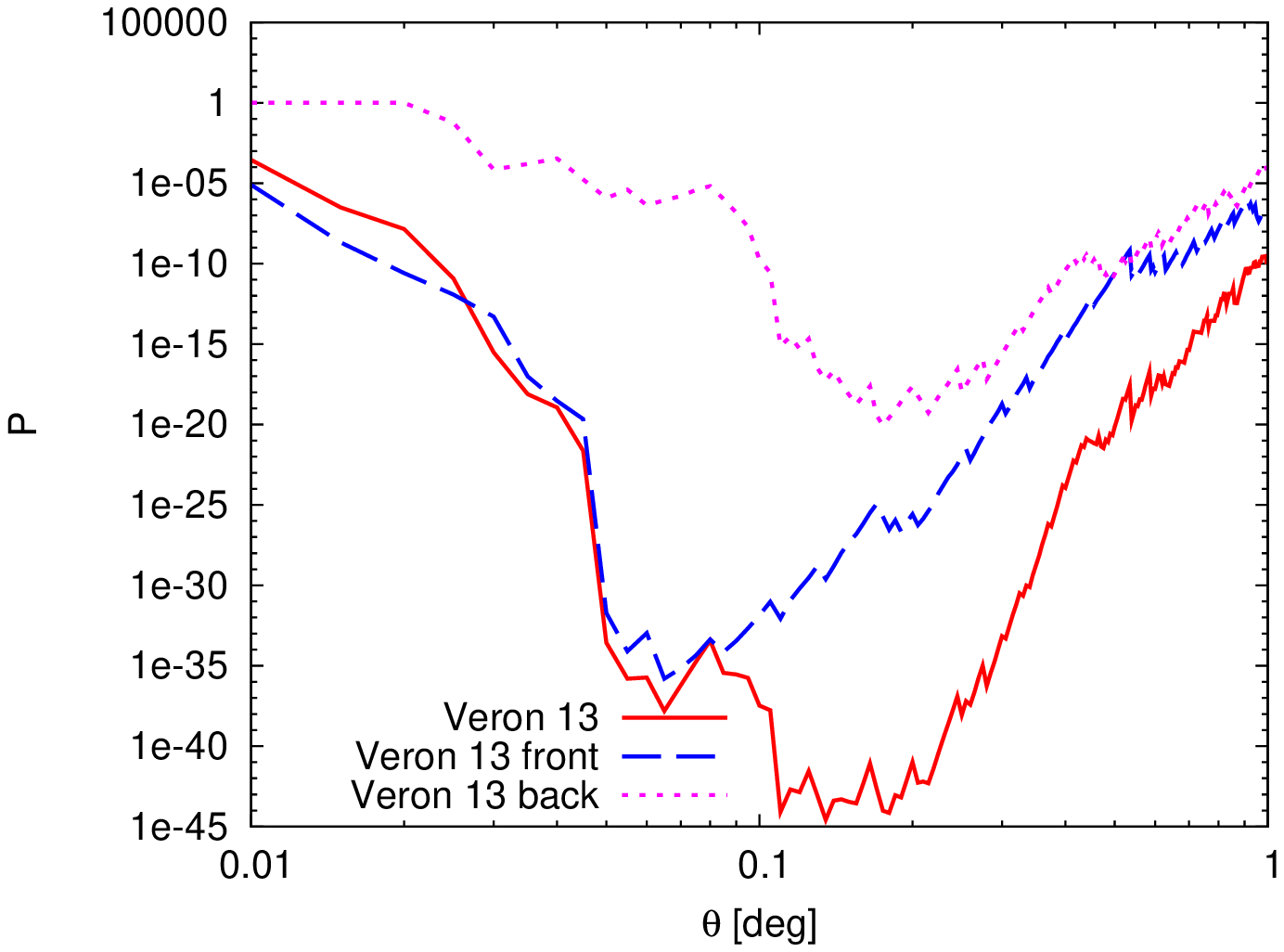}
\end{center}
\caption{Dependence on the chance probability of correlation between arrival directions of $100~\mbox{GeV} \le E  \le 300$~GeV photons and positions of  BL Lacs from  \citet{veron13} catalog as a function of the radius of the search circle. Notations are the same as in Fig. \ref{fig:probability}.}
\label{fig:probabilityV}
\end{figure}
%%%%%%%%%%%%%%%%%%%%%%%%%%%%%%%%%%%%%

%%%%%%%%%%%%%%%%%%%%%%%%%%%%%%%%%%%%%

\begin{figure}[htbp]
\begin{center}
\includegraphics[width=0.7\linewidth,angle=-90]{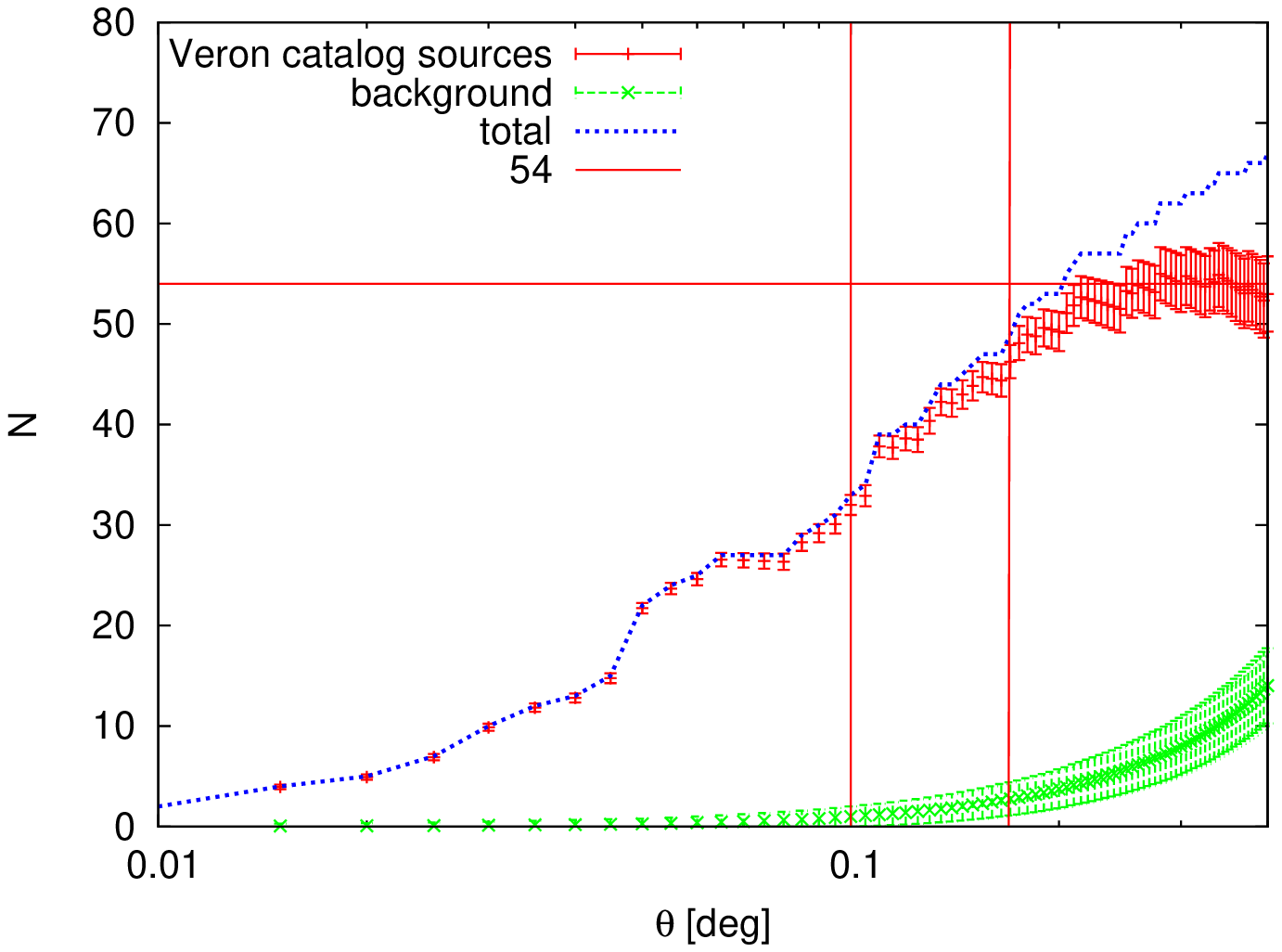}
\end{center}
\caption{Number of events which come from BL Lac sources as a function of the radius of the search circle. Notations are the same as in Fig. \ref{fig:npart}.}
\label{fig:npartV}
\end{figure}
%%%%%%%%%%%%%%%%%%%%%%%%%%%%%%%%%%%%%

Most of the $100~\mbox{GeV} \le E  \le 300$ GeV sources listed in Tables \ref{table1} and \ref{table2} are BL Lac objects. These sources are known to be characterized by relatively hard spectra in the GeV energy band \citep{Fermi_AGN}. It is therefore possible that some of the VHE \gr\ emitting BL Lacs escape detection in the GeV band, while being bright VHE \gr\ sources. 

To search for the VHE \gr\ emitting BL Lacs with very hard spectra, we performed a correlation analysis similar to the one reported above for the first year {\it Fermi} catalog  also for the 13th Veron catalog of BL Lacs~\citep{veron13}. To select a complete sample of BL Lacs, we adopted the prescription of  \citet{10GeV_EGRET} and applied a cut to magnitude of objects $V<18.0$,
which leaves 316 BL Lacs. This set of objects includes six out of seven known VHE \gr\  sources identified in the $E\ge 100$~GeV {\it Fermi} data  by \citet{paper1}.  As in the analysis of the first year {\it Fermi} catalog,   we removed those objects and associated photons from analysis.  We also removed the three photons associated with IC 310. This leaves $N_{\rm source}=310$ BL Lac objects and 
3111 photons. 

The probability that a single photon comes by chance 
near one of the selected BL Lacs is  $p=N_{\rm source} \cdot p_1 = 2.7 \cdot 10^{-4}\cdot \left({\theta}/{0.1^\circ}\right)^2$. The probability that $K$ or more photons fall within an angle $\theta$ from one of the sources is calculated using Eq. \ref{binom}. Fig.~\ref{fig:probabilityV} shows this chance probability as a function of  $\theta$. As for the analysis of first year {\it Fermi} source catalog, the probability has a minimum at the angular resolution of the LAT at the energies $E\ge 100$~GeV. From our correlation analysis we find that 39 BL Lac out of 310 correlate with Fermi data within the angle $\theta\lesssim 0.17^\circ$ corresponding to the minima of $P(\theta)$ from Figs. \ref{fig:probability} and \ref{fig:probabilityV}.

In Fig.~\ref{fig:npartV}, we compare the number of source photons to the number of background photons  as functions of  the angular distance to the BL Lacs. The asymptotic value of the signal events  $N=54$ is reached at the angular distance $\sim 0.2^\circ$.
At the angle $0.1^\circ$, 31 photons come from 28 sources, while only 0.9 of them are on average caused by the background.  At 95 \% C.L.,  there are two background photons.
At the angular distance  $0.17^\circ$, 47 photons come from 39 sources, and 2.6 of them correspond to background in average.  There are 5[6] background photons at the 95\%[99\%] confidence level.
Eight of these 35 sources are not present in Tables \ref{table1} and \ref{table2}. These seven sources are listed in Table \ref{tab:BLLac}. Other sources contributing to the correlation signal in both the first year {\it Fermi} and \citet{veron13} catalogs are  marked by an  ``V" in Tables \ref{table1} and \ref{table2}.

We also performed a special analysis with only front photons and a BL Lac catalog.
In this case, the correlation signal is maximal at $\theta\simeq 0.07^\circ$ (see  Fig.~\ref{fig:probabilityV}). At this angle, there are 21 forward photons,
which correlate with 20 BL Lac sources. PKS 0301-243 has two photons, which can occur by chance in only  $P=10^{-7}$ cases. This means that this source is discovered  by Fermi.
Only 0.18 background photons are expected at this angle, which means that all of the  19 remaining  sources are most probably real. At the 99\% confidence level, two of the 20 sources might be false detections. Sixteen of these 20 sources  are marked as  ``Vf" in  Tables \ref{table1} (15 sources)  and   \ref{table2} (1 source)  and the remaining six are those  have front photons marked ``f" in Table \ref{tab:BLLac}. 

%%%%%%%%%%%%%%%%%%%%%%%%%%%%%%%%%%%%%%%%%%%%
\begin{table}
\begin{tabular}{|l|l|l|l|l|l|l|}
\hline
&Name &              RA &      Dec  & $N_{0.1}$              &$z$\\
& &             deg &      deg  & [$N_{min}$]              &\\
\hline 
1 &{\bf NPM1G +01.0067} &28.165&1.788&1f& 0.08\\
2 & V Zw 326 &  48.210 & 36.256 &  1b &  0.071 \\
3 & B3 0651+428  & 103.682 & 42.799 & 1f&  0.126\\
4 & \underline{RXS J09130-2103} &138.252&-21.054&  1f[1b]& 0.198 \\
5& RXS J10162+4108 &154.070&   41.137& 1b&0.281\\
6& SDSS J114023.48&175.098 & 15.469 &1f& 0.244\\
&+152809.7 &&&&\\
7 & MS 12218+2452  & 186.101& 24.607&  1f& 0.218 \\
8 & MS 13121-4221 & 198.764 & -42.614 &  1f&  0.105 \\

\hline
\end{tabular}
\caption{List of BL Lac sources from \cite{veron13} catalog for which a $100~\mbox{GeV} \le E  \le 300$~GeV  photon is found within the  circle of radius $0.1^\circ$ and are not the first year
{\it Fermi} catalog sources. } \tablefoot{Underlined source is detected by {\it Fermi} in 100-300 GeV band with $TS>25$. Column 5 gives the number of photons in the circles of the radius $0.1^\circ$ and $0.17^\circ$ around the source ($N_{0.1}$ and  $N_{min}$, respectively). Column 6 is the source redshift. }
\label{tab:BLLac}
\end{table}
%%%%%%%%%%%%%%%%%%%%%%%%%%%%%%%%%%%%%%%%%

%%%%%%%%%%%%%%%%%%%%%%%%%%%%%%%%%%%%%
\section{Sources at energies $E>300$ GeV}
\label{sec:300GeV}
%%%%%%%%%%%%%%%%%%%%%%%%%%%%%%%%%%%%%

We discuss correlations between 900 {\it Fermi} photons with $E > 300$ GeV and  sources from the first year {\it Fermi} catalog and BL Lacs   from \cite{veron13} catalog.
We note that this energy is exactly in the range of sensitivity of the TeV gamma-ray telescopes.

%%%%%%%%%%%%%%%%%%%%%%%%%%%%%%%%%%%%%%%%%%%%
\begin{table}[h]
\begin{tabular}{|l|l|l|l|l|l|l|}
\hline
&Name &              RA &      Dec & $N_{0.1}$  & $N_{min}$              &$z$\\
\hline 
1 & {\bf \underline{1ES 0502+675}}  & 76.99 &  67.64  &1f 1b &  1b&  0.341\\
2 & {\bf \underline{Mkn 421}}  &166.1 &  38.21&        1f  3b & 1f & 0.03 \\
3& {\bf \underline{B2 1218+30}} &185.3 & 30.18& 1f&  1f &0.184\\
4& {\bf 1ES 1959+650} &300.0 &  65.13 &1b&  & 0.048\\
\hline
5 & RGB J0250+172 &  42.66 &   17.20 &  &  1b & 1.1  \\
\hline
\hline
6&  SHBL J10102-3119  & 152.6& -31.32&  1b&  & 0.143\\
\hline
\end{tabular}
\caption{List of {\it Fermi} and BL Lac sources  from \cite{veron13} catalog for which a $E  > 300$~GeV photon is found within the  circle of radius $0.17^\circ$.}\tablefoot{ Underlined sources are detected with Fermi at $E>300$ GeV energies with $TS>25$. Notations are the same as in Table \ref{tab:BLLac}.}
\label{tab:300}
\end{table}
%%%%%%%%%%%%%%%%%%%%%%%%%%%%%%%%%%%%%%%%%

We first consider correlations with sources from the first year {\it Fermi} catalog. Using an analysis identical to the one presented in Section \ref{sec:correlation}, we study  correlations of  $E > 300$ GeV 
photons with the sources within a $\theta=0.1^\circ$  angular distance.  There are four sources correlated within this angle, while only 0.7 photons are expected to be found in the direction of {\it Fermi} sources in the case of random coincidences.
Those sources are presented in the first four lines of Table~\ref{tab:300}. All of them are known TeV sources. 

We next consider correlations of photons with $E > 300$ GeV  with sources from the first year {\it Fermi} catalog within  $\theta=0.17^\circ$ in analogy with Table  ~\ref{table2}. There is one more source (source \# 5 in Table ~\ref{tab:300}) that correlates with $E>300$~GeV photons within this angle but does not correlate within $\theta=0.1^\circ$. We expect 2.3 background events within this angle. The  $E>300$~GeV photons from the direction toward source \# 5 is, most probably, a random coincidence, because this source is a blazar at the redshift redshift $z>1$.  Photons with  $E > 300$ GeV cannot reach the Earth from a source at such high redshift because of the interactions with the infrared background. 

We can finally  repeat the analysis of  Section \ref{sec:BL} and consider correlations with BL Lacs   from \cite{veron13} catalog.  We find that there are photons correlating with four out of 
 317 bright BL Lac sources with magnitude  $V<18$ within  $\theta=0.1^\circ$. Those are sources  2, 3, 4, and 6 in Table~\ref{tab:300}. In this case, only 0.27 background photons 
  are on average expected.
 The only new source is source number 6, which is more probably a real source rather than a random coincidence (with probabilities 0.73 vs. 0.27 for the two cases). At the same time, the source is located at a moderate Galactic latitude,  $20^\circ$, so that the possibility that an excess at the position of the source is due to the diffuse emission from the Galaxy cannot ruled out.
 
%%%%%%%%%%%%%%%%%%%%%%%%%%%%%%%%%%%%%
\section{Comments on individual sources}
\label{sec:individual}
%%%%%%%%%%%%%%%%%%%%%%%%%%%%%%%%%%%%%

\subsection{High redshift sources}

Several sources in Table \ref{table1} have redshifts higher than the redshift of 3C 279, $z\simeq 0.5$, which is the furtherest known VHE \gr\ source \citep{3c279_magic}. The study of high redshift sources in the VHE band is interesting for the measurements of the density and cosmological evolution of the extragalactic background light (EBL) \citep{gould,kneiske04,stecker06,mazin07,franceschini08} and the cosmological magnetic fields \citep{neronov09,neronov10c}. In this subsection, we provide some details about the high redshift sources from Table \ref{table1}.
\vskip0.3cm
\noindent\textit{RBS 76} at  $z=0.61$ has two photons within the distance $0.17^\circ$ from the catalog source position. It is detected with a of significance more than $4\sigma$ in the 100-300~GeV band.  The source is also detected with a significance of more than $7\sigma$ in the 30-100~GeV energy band. 

%%%%%%%%%%%%%%%%%%%%%%%%%%%%%%%%%%%%%
\begin{figure}
\includegraphics[width=\linewidth]{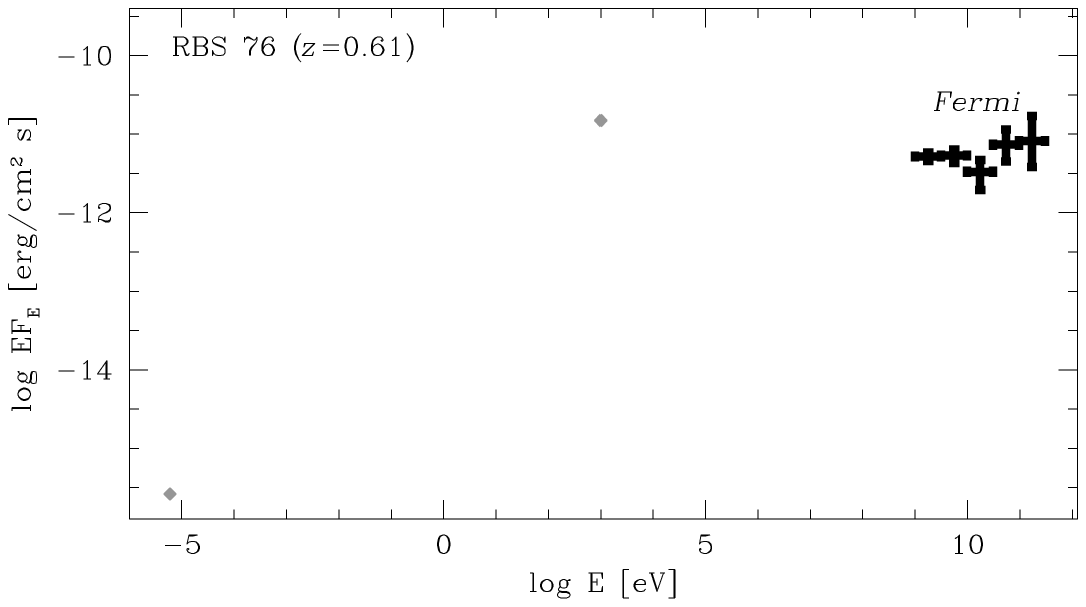}
\caption{Broad-band spectrum of RBS 76. Grey data points are historical data from NED.}
\label{fig:RBS76}
\end{figure}
%%%%%%%%%%%%%%%%%%%%%%%%%%%%%%%%%%%%%

Broad band spectrum of the source is shown in Fig. \ref{fig:RBS76}. The historical multiwavelength data are taken from the NASA Extragalactic Database (NED) \footnote{\tt http://nedwww.ipac.caltech.edu/}.  

\begin{figure}
\includegraphics[width=\linewidth]{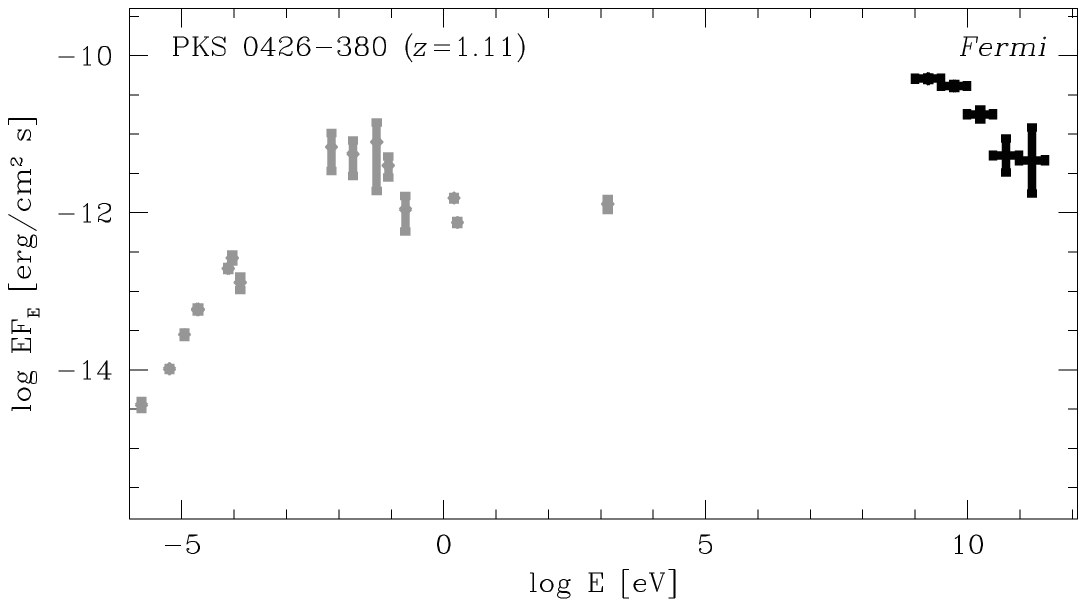}
\caption{Broad band spectrum of PKS 0426-380. Grey data points show historical data taken from NED.}
\label{fig:0426}
\end{figure}

\vskip0.3cm
\noindent\textit{PKS 0426-380} at $z=1.11$ is a bright source in the GeV energy band, with the flux reaching $10^{-10}$~erg/cm$^2$s. Its \gr\ spectrum is relatively soft and  consistent with a power-law with photon index $\Gamma\ge 2.5$. The \gr\ measurements are compared to the multiwavelength data for the source  in Fig. \ref{fig:0426}. As in the case of RBS 76, the overall spectrum of the source could be interpreted in terms of the synchrotron self-Compton or synchrotron external-Compton type models where the radio-to-optical emission is produced by means of the synchrotron mechanism and X-ray-to-\gr\ emission by  inverse Compton scattering. The extremely high redshift of the source and possible presence of the high-energy cut-off in the spectrum might increase the difficulty of the source detection with ground-based \gr\ telescopes  at  energies significantly higher than 100~GeV.

 %%%%%%%%%%%%%%%%%%%%%%%%%%%%%%%%%%%%%
%%%%%%%%%%%%%%%%%%%%%%%%%%%%%%%%%%%%%

\begin{figure}
 \includegraphics[width=\linewidth]{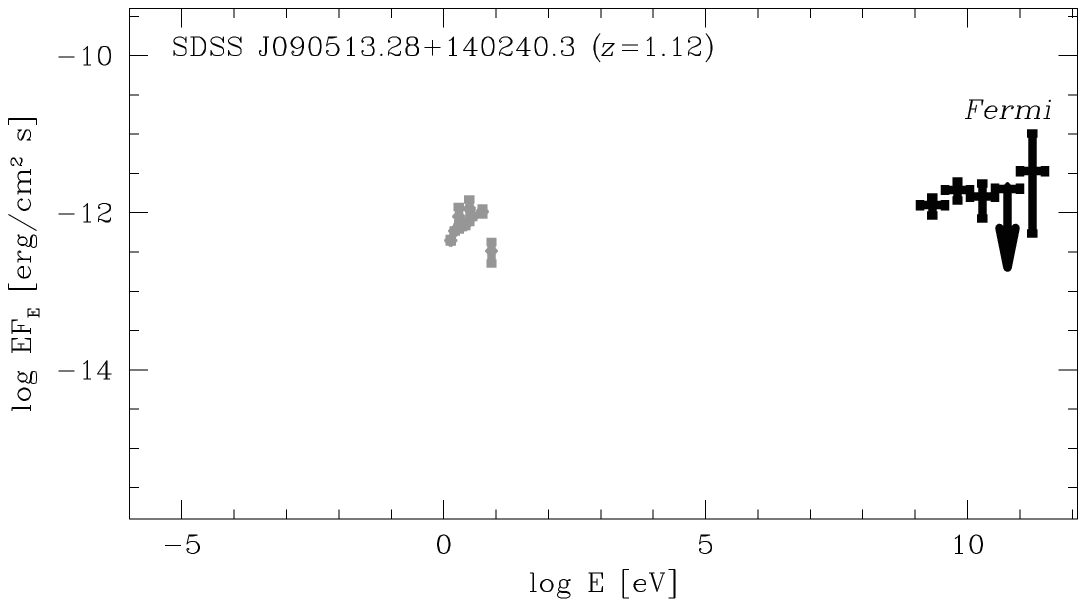}
 \caption{Broad-band spectrum of SDSS J090513.28+140240.3. Grey data points show historical data from NED.}
 \label{fig:SDSS_0905}
 \end{figure}

\vskip0.3cm
\noindent\textit{SDSS J090513.28+140240.3} at $z=1.12$ has a flat ($E^{-2}$) spectrum in the 1-300~GeV band. It was previously only observed in the optical domain. A comparison of \gr\ measurements and historical data is presented in Fig. \ref{fig:SDSS_0905}. The extrapolation of the source spectrum from energies below 100 GeV agrees well with the estimated flux in the 100-300 GeV band, giving no indication of a cutoff. Despite the low statistics in the data for this band obtained with \textit{Fermi/LAT}, this means that the source should be observable with ground-based Cherenkov telescopes.

 %%%%%%%%%%%%%%%%%%%%%%%%%%%%%%%%%%%%%
%%%%%%%%%%%%%%%%%%%%%%%%%%%%%%%%%%%%%

\begin{figure}
\includegraphics[width=\linewidth]{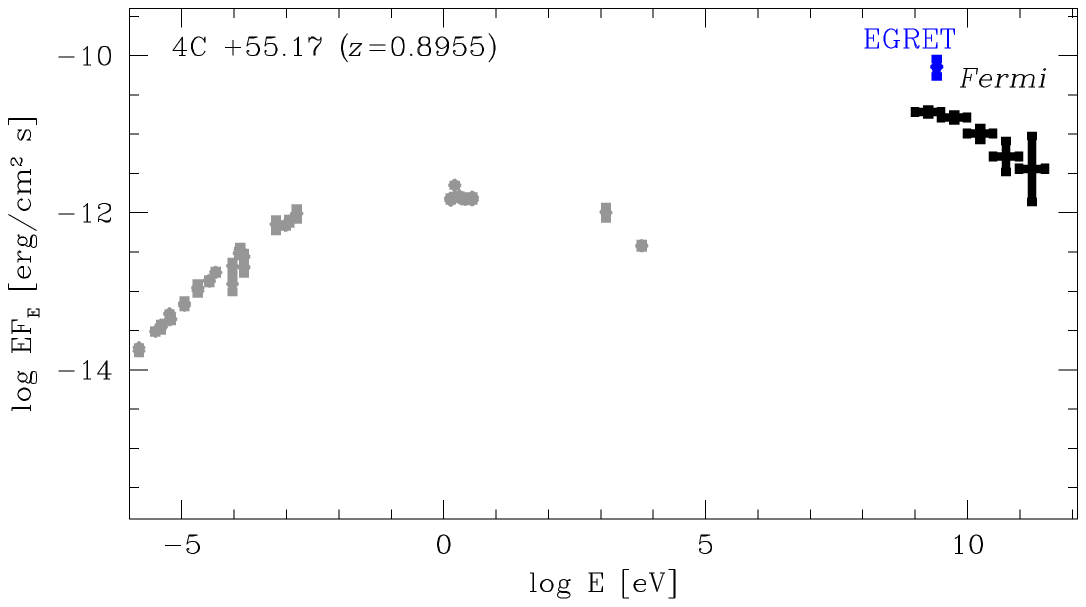}
\caption{Broad-band spectrum of 4C +55.17. Grey data points show historical data from NED. Blue data point is the measurement of \gr\ flux by EGRET \citep{4C55_egret}. }
\label{fig:4C55}
\end{figure}

\vskip0.3cm
\noindent\textit{4C +55.17} at $z=0.8955$ also has a relatively soft spectrum in the 1-300~GeV band. The \gr\ measurements are compared to the broad-band data in Fig. \ref{fig:4C55}. The source was previously detected in the 0.1-1~GeV band by EGRET. There are no other sources from the first year Fermi catalog within a distance $4^\circ$ of the source. This means that an EGRET measurement most probably refers to the same blazar, rather than to a possible nearby higher flux source. A comparison of the historical EGRET measurement with the {\it Fermi} measurement of the source flux shows that the source is variable on a 10~yr timescale.

 %%%%%%%%%%%%%%%%%%%%%%%%%%%%%%%%%%%%%
%%%%%%%%%%%%%%%%%%%%%%%%%%%%%%%%%%%%%

\begin{figure}
\includegraphics[width=\linewidth]{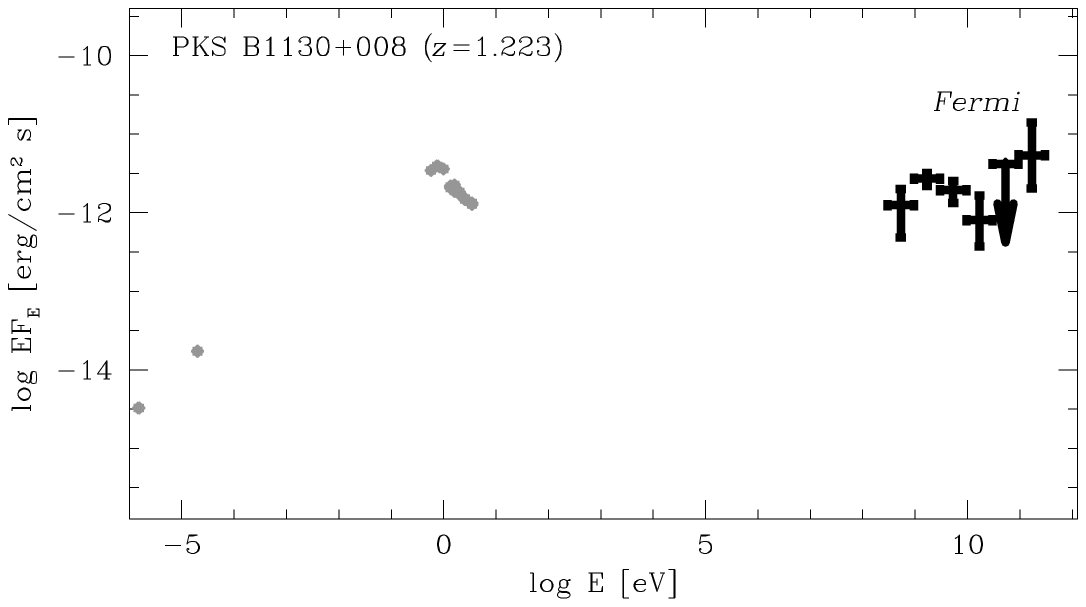}
\caption{Broad-band spectrum of PKS B1130+008.  Grey data points show historical data from NED. }
\label{fig:PKS_B1130}
\end{figure}

\vskip0.3cm
\noindent\textit{PKS B1130+008} at $z=1.223$ has the second highest redshift among the sources listed in Table \ref{table1}. It has an associated front-converted photon with energy 128 GeV. An estimate of the source flux above 100 GeV based on this one photon is above the extrapolation of the source spectrum from the lower energies; we refer to Fig. \ref{fig:PKS_B1130} which possibly indicates that this high redshift source is a false detection. Otherwise, if the source spectrum is flat at energies up to 100 GeV, as indicated by the detection of the $E>100$~GeV \gr , its observation with the ground-based Cherenkov telescopes should provide valuable constraints on the EBL at redshifts $z\sim 1$.

 %%%%%%%%%%%%%%%%%%%%%%%%%%%%%%%%%%%%%
%%%%%%%%%%%%%%%%%%%%%%%%%%%%%%%%%%%%%

\begin{figure}
\includegraphics[width=\linewidth]{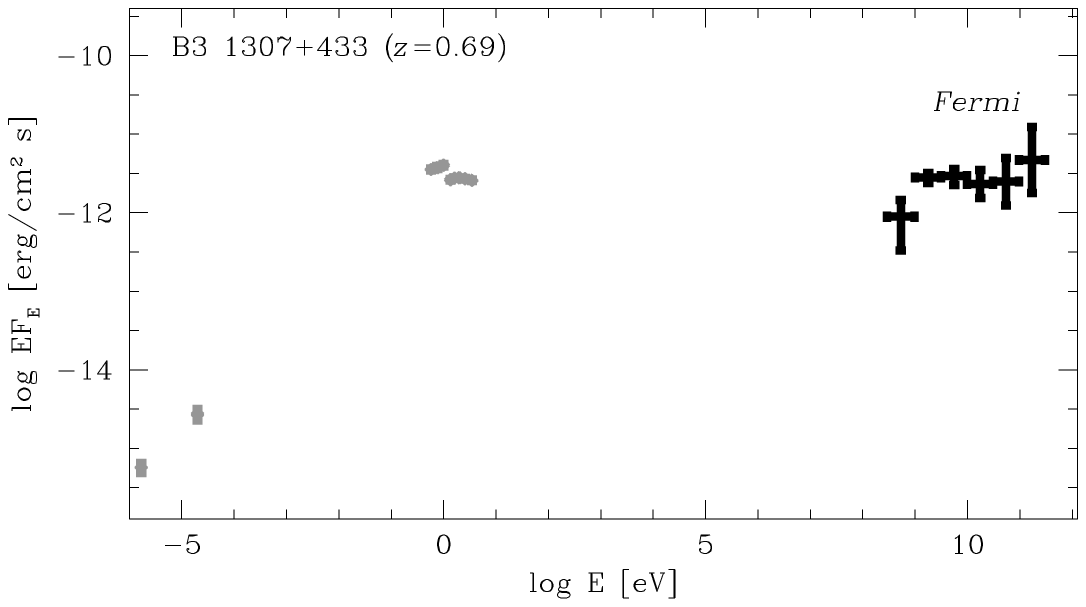}
\caption{Broad-band spectrum of B3 1307+433.  Grey data points indicate historical data from NED. }
\label{fig:B3_1307}
\end{figure}

\vskip0.3cm
\noindent\textit{B3 1307+433} at $z=0.69$ has a harder than $E^{-2}$ spectrum up to the highest energies accessible to {\it Fermi} (see Fig. \ref{fig:B3_1307}). The estimate of the flux based on the detected $E>100$~GeV photon agrees well with the extrapolation of the source spectrum from lower energies. This indicates that the source should be readily detectable with  ground-based \gr\ telescopes, despite its high redshift.

 %%%%%%%%%%%%%%%%%%%%%%%%%%%%%%%%%%%%%
%%%%%%%%%%%%%%%%%%%%%%%%%%%%%%%%%%%%%

\begin{figure}
\includegraphics[width=\linewidth]{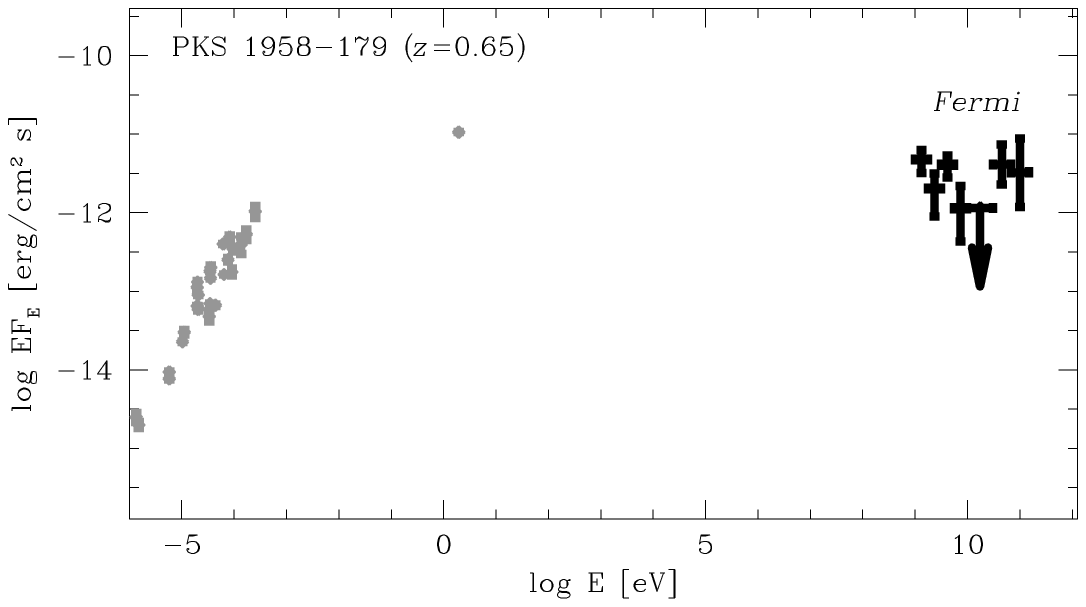}
\caption{Broad-band spectrum of PKS1958-179.  Grey data points show historical data from NED. }
\label{fig:PKS1958}
\end{figure}

\begin{figure}
\includegraphics[width=\linewidth]{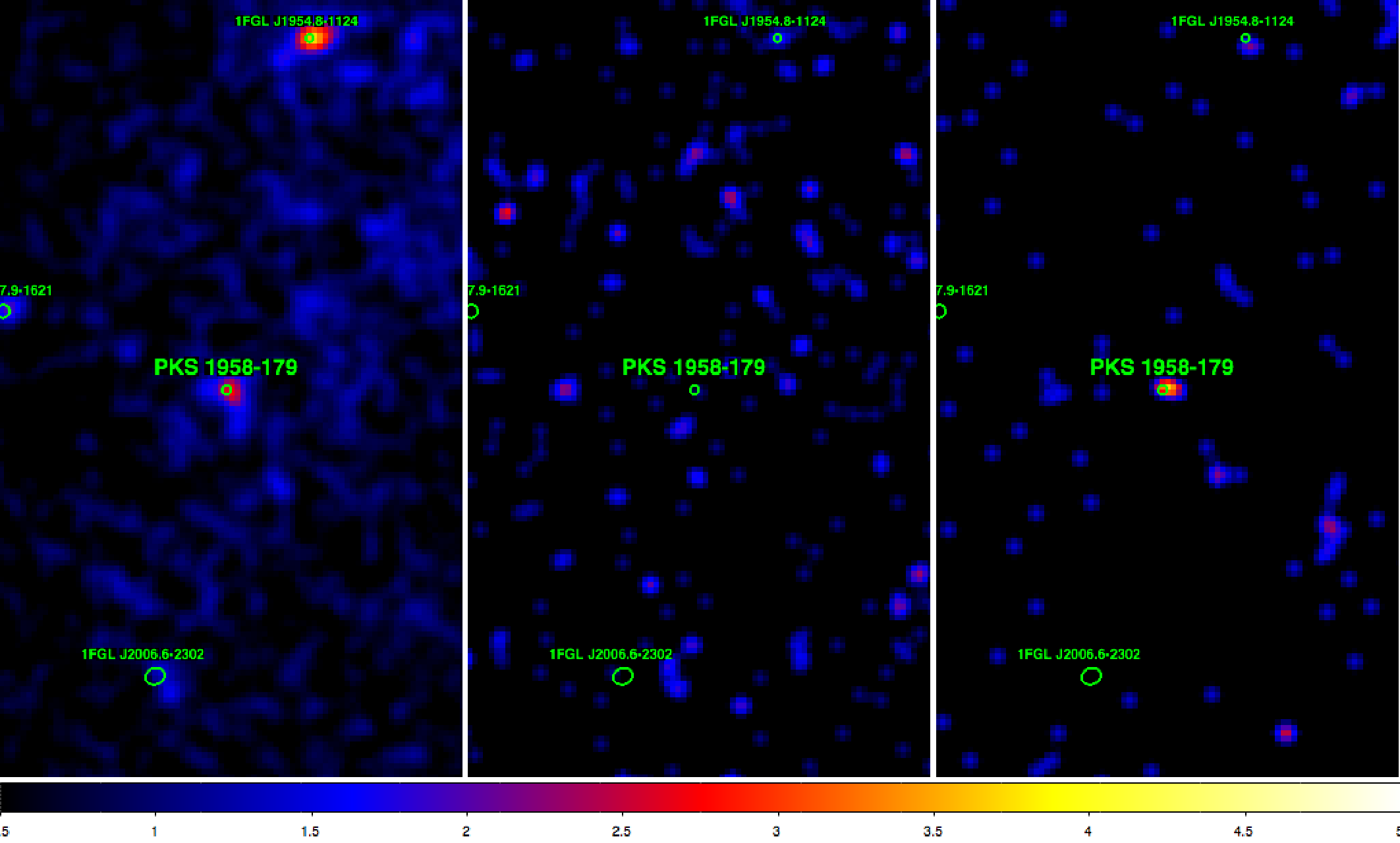}
\caption{LAT count maps smoothed by $0.3^\circ$ around the position of PKS 1958-179 in 1-10 GeV (left), 10-30 GeV (center), and 30-300 GeV (right) bands.  Color scales are linear, from 0.5 to 5 counts per pixel (left panel), from 0.04 to 0.4 cts/pixel (center), and from 0.03 to 0.3 cts/pixel (right panel).}
\label{fig:PKS1958_image}
\end{figure}

\vskip0.3cm
\noindent\textit{PKS 1958-179} at $z=0.65$ is also characterized by an excess of high-energy ($E>30$~GeV) emission above the extrapolation of the lower energy spectrum with $\Gamma=2.46\pm 0.07$ reported in the first year Fermi catalog. No photons are indeed indeed detected from the source in the energy range 10-30~GeV (see Fig. \ref{fig:PKS1958}). However, at the energies above 30 GeV the source is clearly identifiable in Fermi count maps (see Fig. \ref{fig:PKS1958_image}). Standard likelihood analysis gives TS value  $TS=32$, which formally corresponds to $>5 \sigma$ detection in this energy range. This indicates that the source spectrum hardens at the energies above 30~GeV. Unfortunately, the statistic of the Fermi signal at the highest energies is insufficient to clarify the nature of the 100 GeV excess in the source spectrum. 

 %%%%%%%%%%%%%%%%%%%%%%%%%%%%%%%%%%%%%
%%%%%%%%%%%%%%%%%%%%%%%%%%%%%%%%%%%%%

 %%%%%%%%%%%%%%%%%%%%%%%%%%%%%%%%%%%%%
%%%%%%%%%%%%%%%%%%%%%%%%%%%%%%%%%%%%%

\begin{figure}
\includegraphics[width=\linewidth]{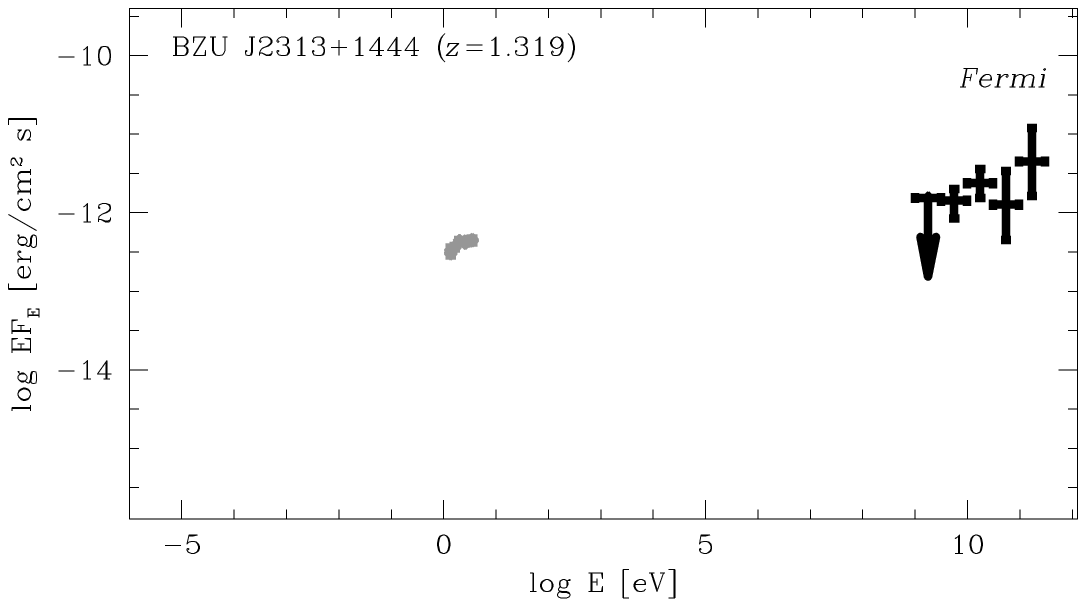}
\caption{Broad-band spectrum of BZU J2313+1444. Grey data points show data from NED. }
\label{fig:BZU}
\end{figure}

\vskip0.3cm
\noindent\textit{BZU J2313+1444} is formally the highest redshift source in Table \ref{table1}, with  $z=1.319$. The source  is relatively weak. In contrast  to the other high-redshift sources, the significance of source detection in the 30-100~GeV band is less than $5\sigma$. From Fig. \ref{fig:BZU}, one can see that the source spectrum does not sxhibit a signature of a high energy cut-off. This implies that the source might be an ideal candidate for the study of EBL evolution with redshift.

 %%%%%%%%%%%%%%%%%%%%%%%%%%%%%%%%%%%%%
%%%%%%%%%%%%%%%%%%%%%%%%%%%%%%%%%%%%%

\subsection{Sources detected at $\ge 4\sigma$ level at $E\ge 100$~GeV}

Apart from the eight high-confidence sources at $E\ge 100$~GeV found by \citet{paper1} and the bright high-redshift sources discussed above, several other sources in Tables \ref{table1} and \ref{table2} have detection significances higher than $4\sigma$ above 100~GeV. We list below these sources and discuss some details of their broad-band spectral energy distribution.

\begin{figure}
\includegraphics[width=\linewidth]{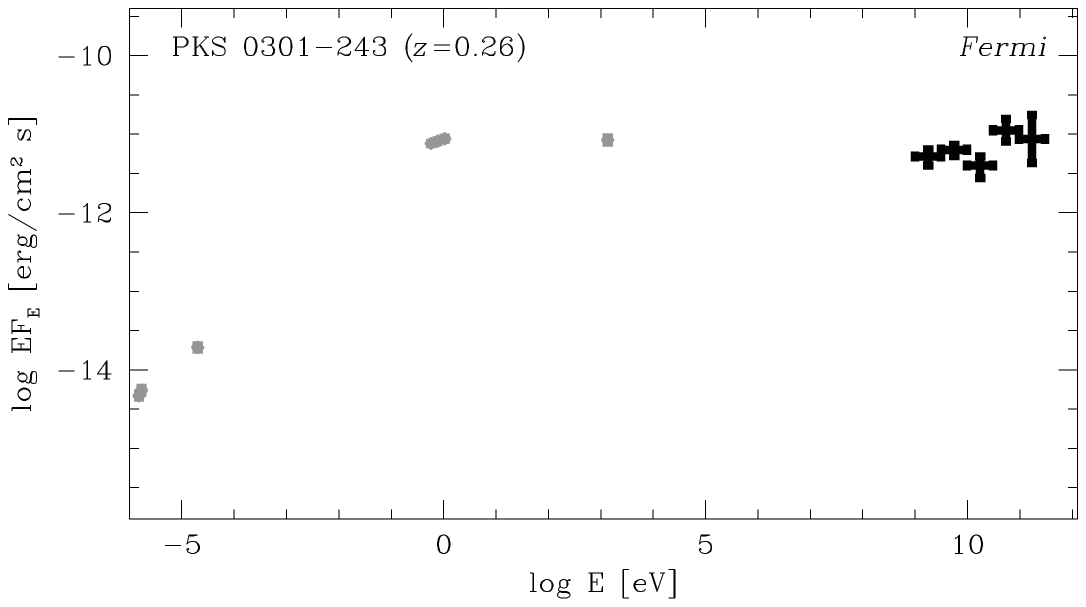}
\caption{Broad-band spectrum of PKS 0301-243. Grey color shows historical data from NED.}
\label{fig:0301}
\end{figure}

\vskip0.3cm
\textit{PKS 0301-243} at $z=0.26$  has one back- and two front-converted photon associated with the source above 100~GeV. The back photon is within the 95\% containment circle of the radius $\theta=0.3^\circ$.  Taken together, the two photons represent a $\simeq 5.2\sigma$ detection of the source above 100~GeV. At the same time, the source is also detected with high significance in the 30-100~GeV band. 
Both $E>100$~GeV photons came during the flaring activity of the source in April-May 2010 \citep{PKS0301_Fermi,PKS0301_NSV}.
The broad-band spectrum of PKS 0301-243, shown in Fig. \ref{fig:0301}, is consistent with the possibility that radio-to-X-ray emission are produced via the synchrotron mechanism and the 1-300~GeV emission via the inverse Compton mechanism.

%%%%%%%%%%%%%%%%%%%%%%%%%%%%%%%%%%%%%
%%%%%%%%%%%%%%%%%%%%%%%%%%%%%%%%%%%%%

\begin{figure}
\includegraphics[width=\linewidth]{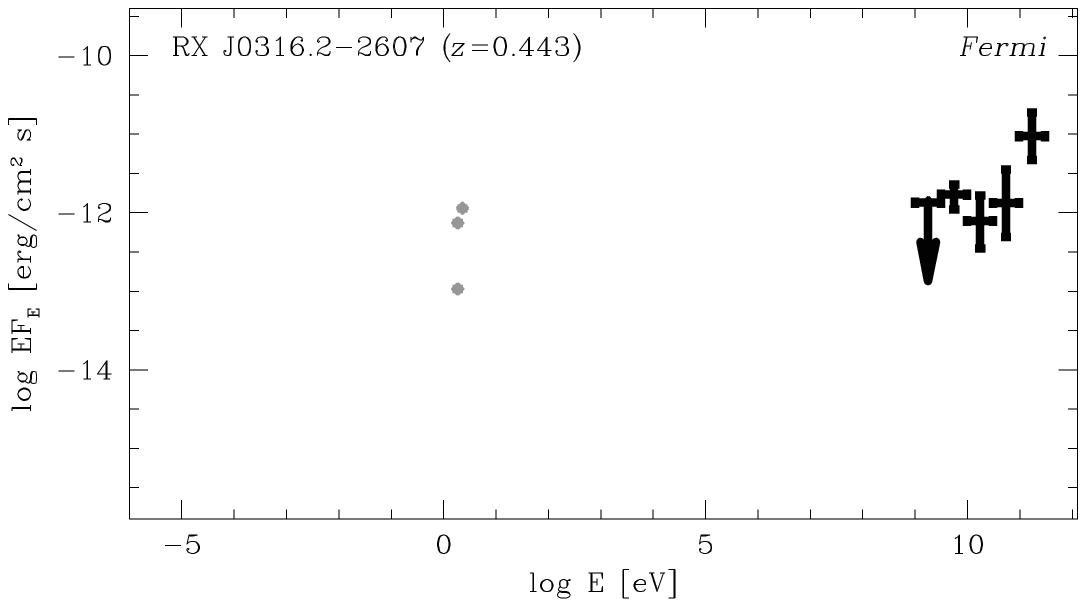}
\caption{Broad-band spectrum of RX J0316.2-2607. Grey color shows historical data from NED.}
\label{fig:0316}
\end{figure}

\vskip0.3cm
\noindent\textit{RX J0316.2-2607} is a relatively high redshift ($z=0.443$) source, which has two back \gr s within the distance $\theta=0.1^\circ$.  The chance coincidence probability of finding two photons within this distance from the source is $2.5\times 10^{-6}$, which implies a detection significance $4.7\sigma$ above 100~GeV. From Fig. \ref{fig:0316}, one can see that the source has a hard \gr\ spectrum without a signature of a high-energy cut-off. The \gr\ energy flux from the source is comparable to the optical flux. Similarly to BZU J2313+1444, this source might also be an ideal candidate for the study of both EBL and  cosmological magnetic fields.

 %%%%%%%%%%%%%%%%%%%%%%%%%%%%%%%%%%%%%
%%%%%%%%%%%%%%%%%%%%%%%%%%%%%%%%%%%%%

\begin{figure}
\includegraphics[width=\linewidth]{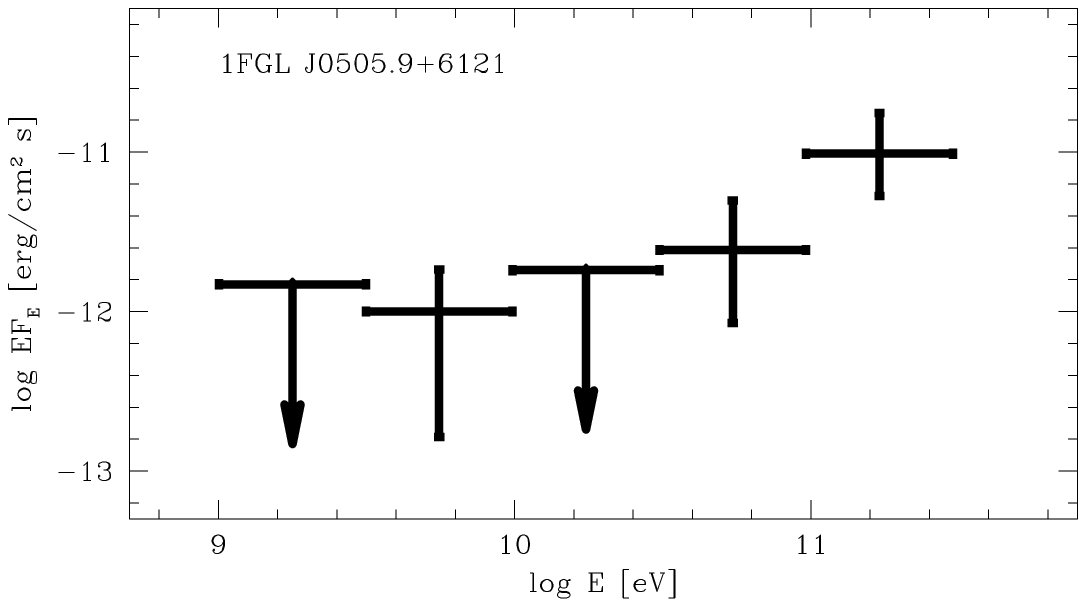}
\caption{{\it Fermi} spectrum of 1FGL J0505.9+6121.}
\label{fig:0505}
\end{figure}

\vskip0.3cm
\noindent\textit{1FGL~ J0505.9+6121} has three adjacent photons within $0.22^\circ$. Formally, the chance probability  of finding three photons inside this distance is  $P=8 \cdot 10^{-8}$; however, since the detection angle was not defined a priori, it has to be penalized for this and the final probability is just $5\sigma$ (see Table \ref{table1}), still implying that the source is detected at $E\ge 100$~GeV  individually, and not only within the population of {\it Fermi} sources from Table \ref{table1}. It is also detected with a significance higher than $5\sigma$ in the 30-100~GeV band. The spectrum of the source shown in Fig. \ref{fig:0505} is consistent with a hard power-law (photon index harder than 2) up to the highest energy.

 %%%%%%%%%%%%%%%%%%%%%%%%%%%%%%%%%%%%%
%%%%%%%%%%%%%%%%%%%%%%%%%%%%%%%%%%%%%

The object is located near the Galactic plane at $b=+12^\circ$, so that it can be of Galactic origin. The high level of Galactic diffuse \gr\ background around the location of the source prevents its detection at lower energies $E\le 1$~GeV. The small value of the test-statistics (TS) value \footnote{See http://fermi.gsfc.nasa.gov/ssc/data/analysis/scitools} for this source in the Table \ref{table1} could indicate that this source is not point-like. A standard \textit{Fermi/LAT} analysis procedure, which was used to calculate the value of TS, does not work in this case.

Detailed multiwavelength observations are needed to constrain its nature. A possible candidate for the source identification is the radio- and X-ray loud AGN RX J0505.9+6113 \citep{brinkman97} situated at the distance $7.6'$ from the catalog source position and the distance $0.06^\circ$ from the front-converted $E\ge 100$~GeV photon.
 
 %%%%%%%%%%%%%%%%%%%%%%%%%%%%%%%%%%%%%
\begin{figure}
\includegraphics[width=\linewidth]{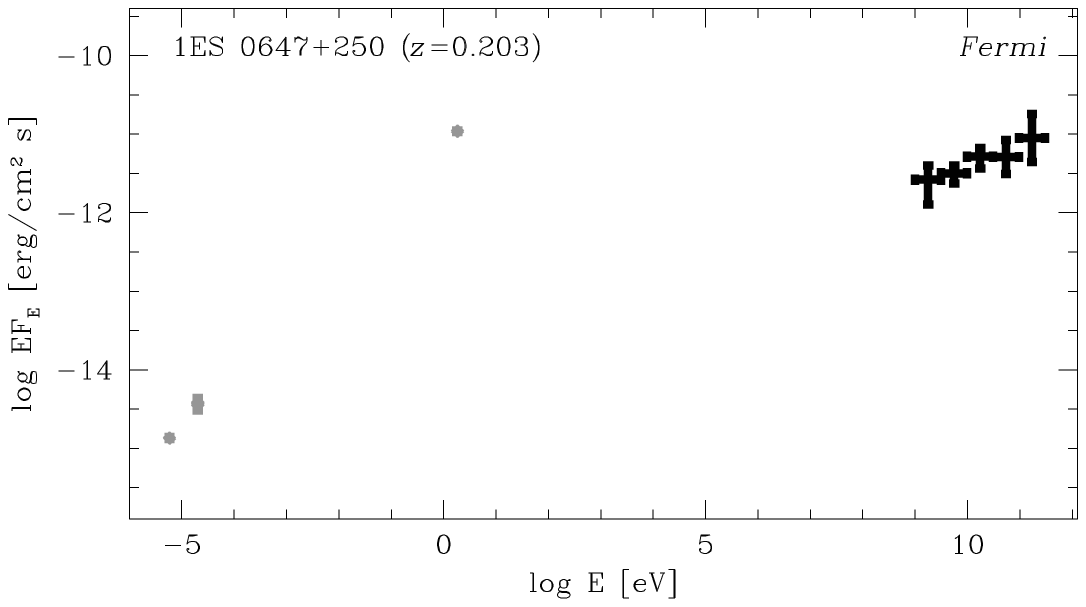}
\caption{Broad-band spectrum of 1ES 0647+250. Grey color shows historical data from NED.}
\label{fig:0647}
\end{figure}
%%%%%%%%%%%%%%%%%%%%%%%%%%%%%%%%%%%%%

\vskip0.3cm
\noindent\textit{1ES 0647+250} at $z=0.201$ was already mentioned as a "candidate" TeV blazar in the analysis of \citet{costamante02}, based on its broad-band spectra  properties. Fig. \ref{fig:0647} demonstrates that the source has a hard \gr\ spectrum, with no signature of a cut-off up to 300~GeV. This ensures that it should be readily detectable with ground-based \gr\ telescopes. 

\begin{figure}
\includegraphics[width=\linewidth]{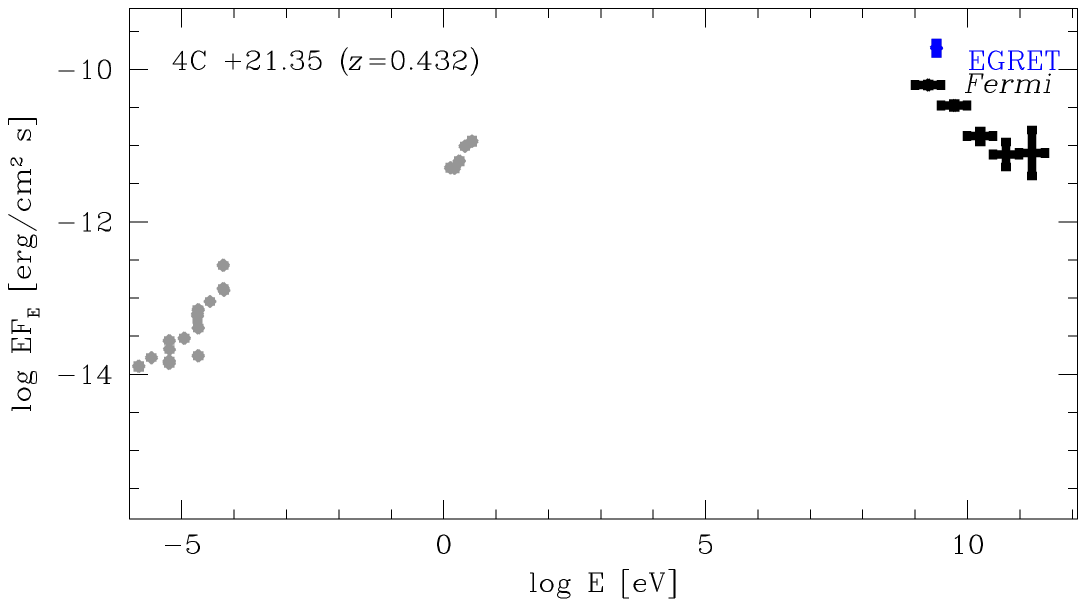}
\caption{Broad-band spectrum of 4C +21.35. Grey color shows historical data from NED. Blue point is EGRET measurement.}
\label{fig:4C+21}
\end{figure}

\vskip0.3cm
\noindent\textit{4C +21.35} at $z=0.432$ is an intermediate redshift BL Lac that exhibited a \gr\ flare in April-May 2010 \citep{4C_Fermi_ATEL} during which both $E>100$~GeV photons associated with the source were detected, resulting in the detection of the source with $5.6\sigma$ significance above 100~GeV \citep{4C_NSV_ATEL}. The detection of the source above 100~GeV was recently confirmed with MAGIC observations of the source \citep{4C_ATEL_MAGIC}. A time-averaged spectrum of the source in the 1-300 GeV band is shown, together with the historical data, in Fig. \ref{fig:4C+21}. The previous EGRET measurement of the source flux is characterized by a somewhat higher source flux. This indicates that the source is variable not only on the short (days to month) timescale, but also on decade-long timescales.

 %%%%%%%%%%%%%%%%%%%%%%%%%%%%%%%%%%%%%
%%%%%%%%%%%%%%%%%%%%%%%%%%%%%%%%%%%%%

\subsection{Sources from the BL Lac catalog}

\begin{figure}
\includegraphics[width=\linewidth]{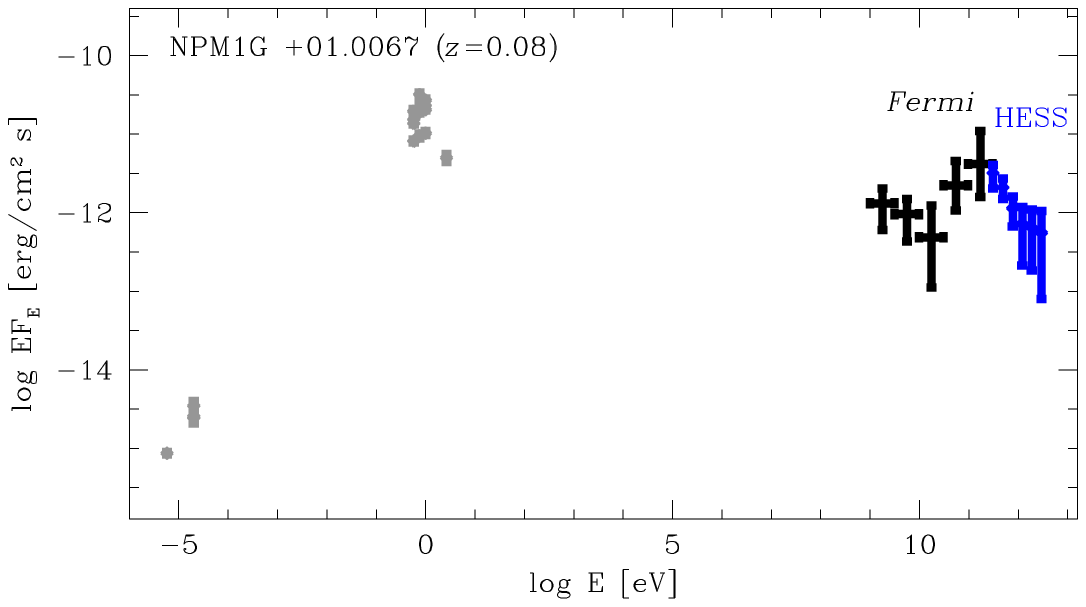}
\caption{Broad-band spectrum of NPM1G +01.0067. Grey color shows historical data from NED. }
\label{fig:NPM}
\end{figure}

{\it NPM1G+01.0067} at $z=0.08$ was previously detected in the VHE band by HESS \citep{NPM_HESS}. 
The {\it Fermi} spectrum of the source, shown in Fig. \ref{fig:NPM} is in good agreement with the HESS measurements. In spite of the source not being reported in the first year {\it Fermi} catalog, it is detected in the 1.9~yr  exposure with a TS value  of $TS=108$ above 1 GeV.

 %%%%%%%%%%%%%%%%%%%%%%%%%%%%%%%%%%%%%
%%%%%%%%%%%%%%%%%%%%%%%%%%%%%%%%%%%%%

{\it RX J09130-2103} has two \gr s with energies $E\ge 100$~GeV within $0.1^\circ$ distance. One of the \gr s is a front-converted \gr, for which the 68\% containment circle has a smaller radius. The front-converted \gr\ is at the distance $0.01^\circ$ from the catalog source  position and a back photon at $0.03^\circ$. The field of the radius $10^\circ$ around the source contains 20 photons at the energies above 100~GeV. The overall chance probability of  finding  front- and back-converted photons within the 68\% containment circles of the front-converted and back-converted photons  is $\simeq 10^{-5}$.

Despite  the source not being  reported in the first year {\it Fermi} catalog,  it is also detected with {\it Fermi} at energies below 100~GeV. A standard likelihood analysis in the energy range 1-300~GeV gives the TS value 80, which corresponds to a source detection  significance $\simeq 9\sigma$. 

\begin{figure}
\includegraphics[width=\linewidth,height=0.9\linewidth]{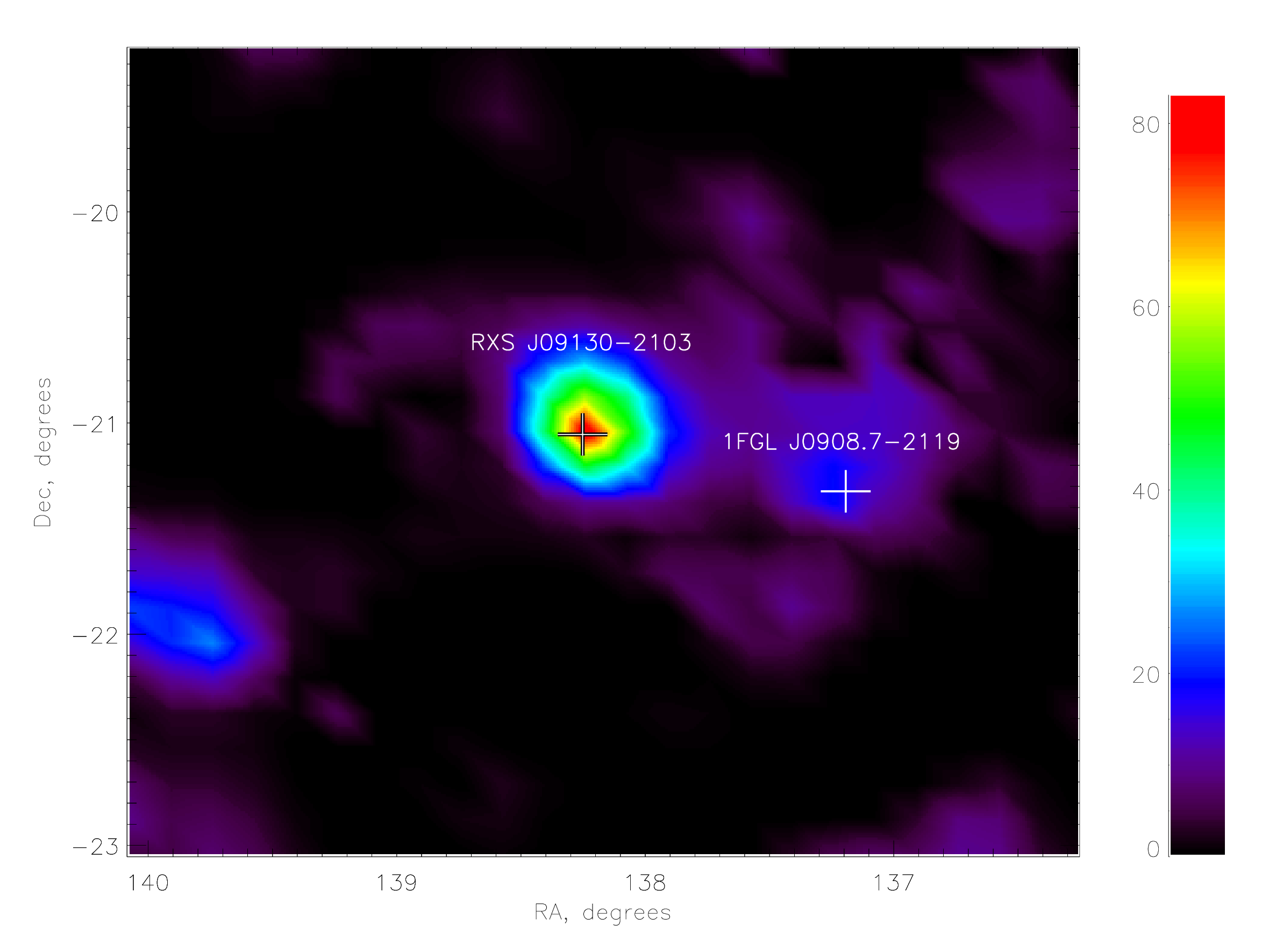}
\caption{TS map of the region $2^\circ\times 2^\circ$ in the energy band above 1~GeV around the position of RX J0913.0-2103. }
\label{fig:TS_RXJ0913}
\end{figure}

The map of TS values in the region $2^\circ\times 2^\circ$ around the source is shown in Fig. \ref{fig:TS_RXJ0913}. The TS values were calculated for the model of source distribution in the region $14^\circ$ around RX J09130-2103 in which all the first year {\it Fermi} source catalog sources apart from  1FGL J0908.7-2119 were included and one additional source was allowed to have a  variable position. One can see that the TS value at the position of RX J09130-2103 is much larger than that at the position of 1FGL J0908.7-2119, which is only marginally detected  above 1~GeV.

\begin{figure}
\includegraphics[width=\linewidth]{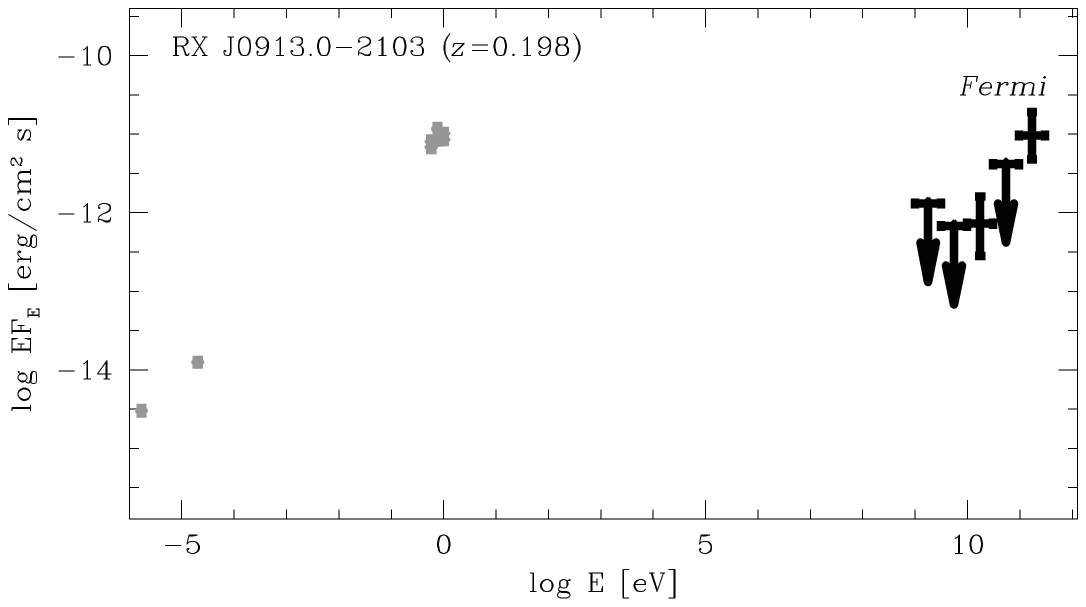}
\caption{Broad-band spectrum of RX J0913.0-2103. Grey data points show historical data from NED. }
\label{fig:RXJ0913}
\end{figure}

 The broad-band spectrum of the source is shown in Fig. \ref{fig:RXJ0913}. One can see that the \gr\ spectrum of the source is hard, which explains its non-detection in the first 11 months of operation of {\it Fermi}. 

%%%%%%%%%%%%%%%%%%%%%%%%%%%%%%%%%%%%%
%%%%%%%%%%%%%%%%%%%%%%%%%%%%%%%%%%%%%

%%%%%%%%%%%%%%%%%%%%%%%%%%%%%%%%%%%%%
%%%%%%%%%%%%%%%%%%%%%%%%%%%%%%%%%%%%%

\vskip0.3cm
{\it MS 12218+2452} has a front-converted  photon at the distance $0.04$ from the catalog source position. The source is also not reported in the first year Fermi catalog, but is detected in the 1.5~yr exposure with TS value 80, which corresponds to the $\simeq 9\sigma$ significance of the source detection. 

\begin{figure}
\includegraphics[width=\linewidth,height=0.9\linewidth]{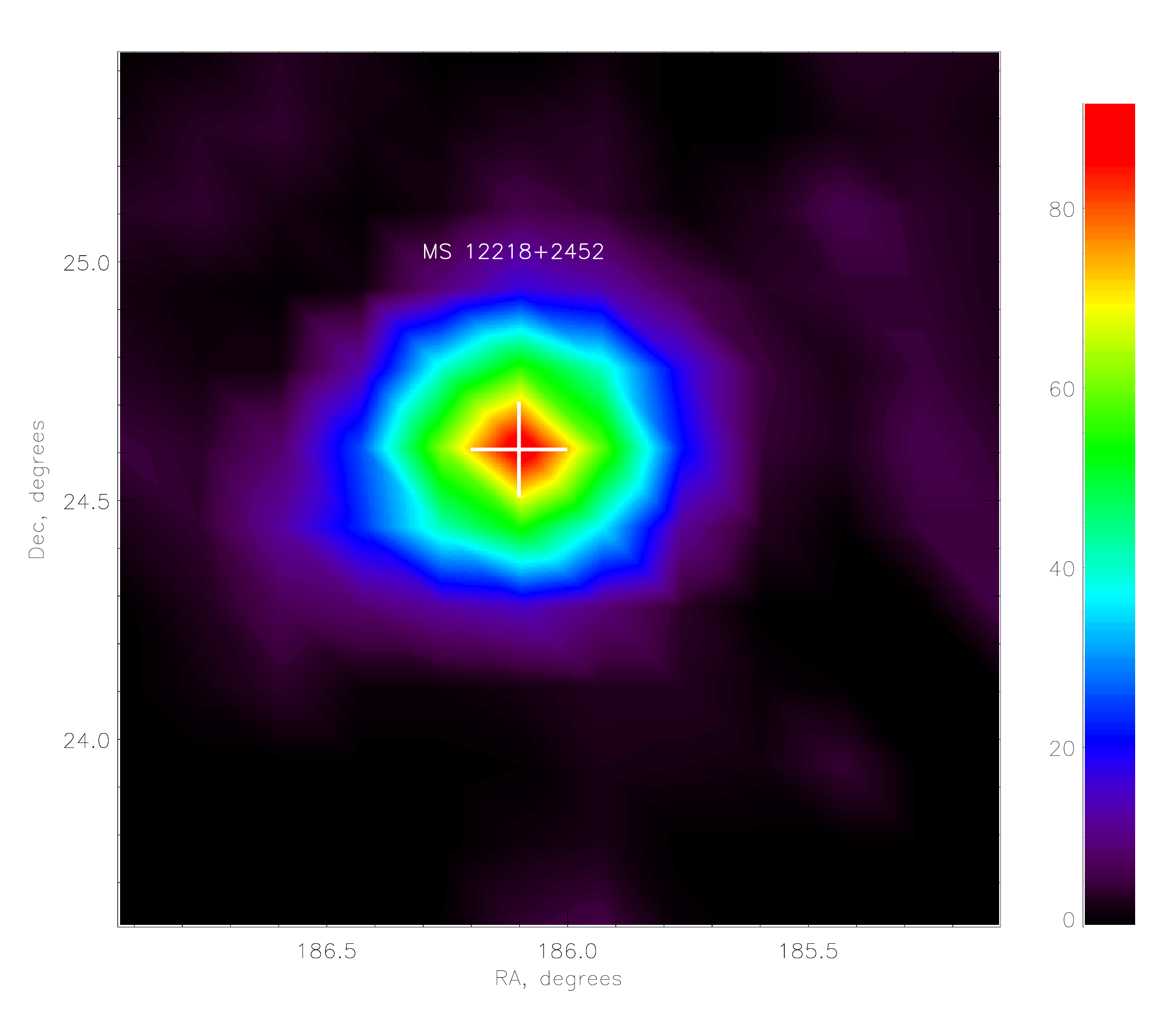}
\caption{TS map of the region $1^\circ\times 1^\circ$ in the energy band above 1~GeV around the position of MS 12218+2452. }
\label{fig:TS_MS1221}
\end{figure}

The map of TS values in the region of size $1^\circ\times 1^\circ$ at  energies $E\ge 1$~GeV around the source position is shown in Fig. \ref{fig:TS_MS1221}. To construct this map, we included all the sources from the first year {\it Fermi} catalog situated at the distance $\theta\le 14^\circ$ from the source position into the model of source distribution on the sky. One can clearly see the source in the TS map.

\begin{figure}
\includegraphics[width=\linewidth]{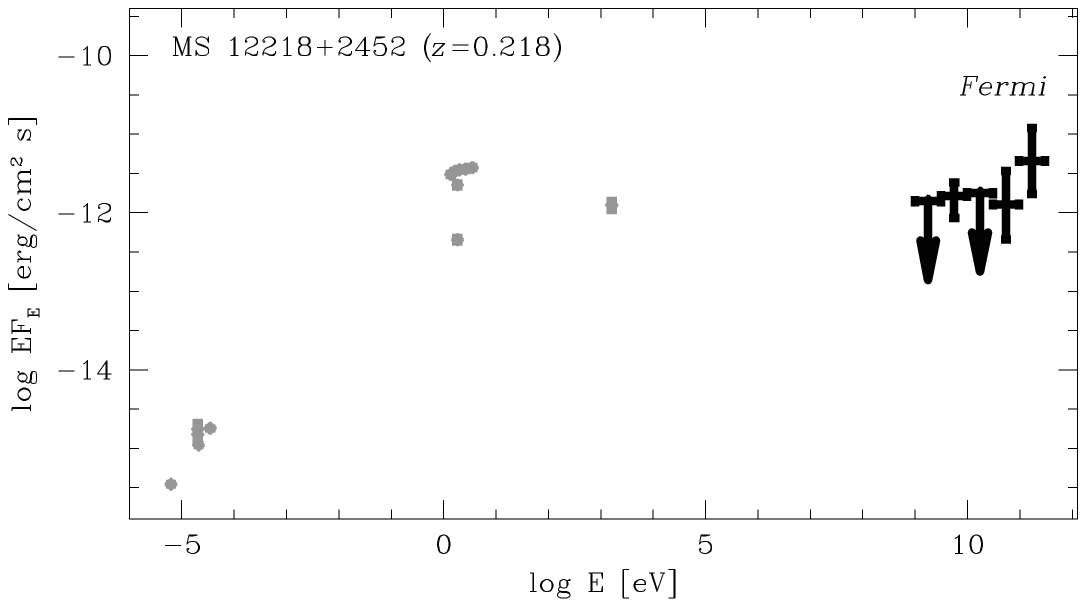}
\caption{Broad-band spectrum of MS 12218+2452. Grey data points show historical data from NED. }
\label{fig:MS1221}
\end{figure}

{\it Fermi} data are compared to the broad-band archival data in Fig. \ref{fig:MS1221}. The \gr\ flux from the source is comparable to its optical flux intensity. The \gr\ spectrum does not display any signatures of a high-energy cut-off. 

%%%%%%%%%%%%%%%%%%%%%%%%%%%%%%%%%%%%%
%%%%%%%%%%%%%%%%%%%%%%%%%%%%%%%%%%%%%

%%%%%%%%%%%%%%%%%%%%%%%%%%%%%%%%%%%%%
%%%%%%%%%%%%%%%%%%%%%%%%%%%%%%%%%%%%%

%%%%%%%%%%%%%%%%%%%%%%%%%%%%%%%%%%%%%
\section{Discussion and conclusions}
\label{sec:discussion}
%%%%%%%%%%%%%%%%%%%%%%%%%%%%%%%%%%%%%

The study of cross-correlation of the arrival directions of the highest energy photons ($E\ge 100$~GeV) detected by {\it Fermi} with the catalog of sources detected at lower energies $E<100$~GeV has proven to be an efficient way of finding new \gr\ sources observable in the VHE \gr\ band. This study complements the pointed observations of selected \gr\ sources with the ground-based \gr\ telescopes because it results in an all-sky survey of the high Galactic latitude VHE \gr\ sky. 

The  high Galactic latitude sky survey above 100~GeV has revealed a large number of new VHE \gr\ sources. Adding the list of  new sources to that of all  sources previously detected with the ground-based \gr\ telescopes doubles the number of known extragalactic VHE \gr\ sources. As for all known VHE \gr\ sources, most of the new sources reported in Tables 1-3 are BL Lac type objects, a special type of AGN with jets closely aligned to the line of sight. 

The majority of the sources listed in Table \ref{table1} are real new VHE \gr\ sources; only three sources are expected to be false detections caused by the chance coincidence of arrival directions of \gr s with the source position. Several sources in Table 1 have more than one photon associated with them. These sources are detected with high significance in the energy range 100-300~GeV. Sources correlating with only one $E>100$~GeV photon should be considered as "VHE source candidates". Apart from the known VHE \gr\ sources, the unidentified {\it Fermi} source 1FGL J0505.9+6121 as well as PKS 0301-243 and 4C +21.35 is detected with a significance close to $5\sigma$  above 100~GeV. In addition, BL Lacs with an associated front-converted photon, which are found to correlate with the \citet{veron13} catalog (marked by "Vf" in Table \ref{table1}), should also be considered as high-confidence detections. 

Table \ref{table2} lists 21 lower significance sources, among which five or six are most probably false detections. As for Table 1, the source marked "Vf" (\# 56) which contributes with a front-converted photon to the correlation with the BL Lac catalog should be considered as a higher significance detection. For the sources listed in this table, it is reasonable to accumulate longer exposure to single out false detections. 

Table 3 lists VHE \gr\ emitting BL Lacs that were not listed in the first year {\it Fermi} catalog, but contribute to the correlation of the {\it Fermi} VHE \gr s with the BL Lac catalog of \citet{veron13}. Their absence from the {\it Fermi} catalog might be caused by the hard spectra of these sources. We have demonstrated that three of the eight sources listed in Table 3, the sources \# 1, 4, and 7 are, in fact, detected with a significance $> 5\sigma$ in the 1-300~GeV band. The other three sources which have a front-converted photon associated with them should also be considered as probable true detections. 

Table 4 lists six sources, which have photons with energies above 300 GeV, including four known TeV sources and  two possible new sources which are, however, consistent with  random coincidences with background photons.  Among the TeV sources, only B2 1218+30 does not 
contain photons in the 100-300 GeV energy range within a region $\theta \le 0.2^\circ$ about its center. However it has 2 back-converted photons at $0.2^\circ < \theta < 0.3^\circ$.

Some of the new VHE \gr\ sources listed in Table \ref{table1} is situated at high redshifts $z\sim 1$. Our analysis of the \gr\ spectra of the high redshift sources indicates that at least several of them do not display any signatures of a high-energy cut-off up to 300~GeV. The typical fluxes of the sources listed in Tables 1-3 are at the level $EF_E(E\ge 100\mbox{ GeV})\ge (few)\times 10^{-12}$~erg/(cm$^2$s), i.e., at the level of 0.1 Crab units. These sources should be readily detectable by  ground-based Cherenkov telescopes at energies of several hundreds of GeV. The larger collection area of the ground-based telescopes should result in much higher quality signal statistics, which should allow a study of the detailed spectral properties of these new bright VHE \gr\ sources. We anticipate that a detailed high sensitivity study of high-redshift VHE \gr\ sources with  ground-based \gr\ telescopes will enable us  to study the cosmological evolution of the BL Lac properties, the EBL, and the associated evolution of galaxies producing the EBL \citep{gould,kneiske04,stecker06,mazin07,franceschini08}, as well as cosmological intergalactic magnetic fields \citep{neronov09,neronov10c}. 

%%%%%%%%%%%%%%%%%%%%%%%%%%%%%%%%%%%%%
\section{Acknowledgments}
We would like to thank Igor Tkachev and Peter Tinyakov, who performed an analysis similar to the one described  in  Section  \ref{sec:BL} and 
pointed out to us that we can also use data with $E>300$ GeV obtained by {\it Fermi}. The work of AN and IeV is supported by the Swiss National Science Foundation grant PP00P2\_123426.  
%%%%%%%%%%%%%%%%%%%%%%%%%%%%%%%%%%%%%

\end{document}